\def\DIRvalue{QiuZabzine}
\def\IDvalue{QZ}
\def\titlevalue{Review of localization for\\
 5d supersymmetric gauge theories}

\def\authorvalue{Jian Qiu$^{a,b,c}$ and Maxim Zabzine$^c$}
\def\shortauthorvalue{Jian Qiu and Maxim Zabzine}

\def\addressvalue{${}^a$Max-Planck-Institut f\"ur Mathematik,\\
 Vivatsgasse 7,
53111 Bonn, Germany\\
      \vspace{.5cm}
   ${}^b$ Department of Mathematics,  Uppsala University, \\
   Box 480, SE-75106 Uppsala, Sweden\\
     \vspace{.5cm}
${}^c$Department of Physics and Astronomy,
     Uppsala university,\\
     Box 516,     SE-75120 Uppsala,     Sweden\\
\tt jian.qiu@math.uu.se, Maxim.Zabzine@physics.uu.se
}

\def\abstractvalue{We give a pedagogical review of the localization of supersymmetric gauge theory on 5d toric Sasaki-Einstein manifolds.
 We construct  the cohomological complex resulting from supersymmetry  and consider its natural toric deformations with all equivariant
  parameters turned on.
We also give detailed discussion on how the Sasaki-Einstein geometry permeates every aspect of the calculation, from Killing spinor, vanishing theorems to the index theorems.
}

\def\preprintvalue{UUITP-09/16}

\ifx\ifLONG\undefined
\documentclass[12pt]{article}
\newcommand{\chapterauthor}[1]{
\begin{center}
{\bf \normalsize  #1}
\end{center}
}

\newcommand{\chapteraddress}[1]{
\begin{center}
{ \small \it \addressvalue}
\end{center}
}

\newcommand{\chapterabstract}[1]{
\vspace{\baselineskip}
\begin{center}
\textbf{\small Abstract}
\end{center}
#1}

\newcommand{\chapterheader}{

\chapter[\titlevalue{}  (by \shortauthorvalue)]{\titlevalue}
\label{Chapter\IDvalue}
\chapterauthor{\authorvalue}
\chapteraddress{\addressvalue}
\chapterabstract{\abstractvalue}
\tightmtctrue
\minitoc
}

\newcommand{\documentheader}{
\begin{flushright} \small
  \preprintvalue
 \end{flushright}

\begin{center}
{\bf \Large \titlevalue}
\end{center}

\chapterauthor{\authorvalue}
\chapteraddress{\addressvalue}
\chapterabstract{\abstractvalue}

\medskip

This is a contribution to the review volume ``Localization techniques
in quantum field theories'' (eds. V.~Pestun and M.~Zabzine) which
contains 17 Chapters available at \cite{ContributionSummary}

\tableofcontents
}

\newcommand{\ifvolume}[2]{\ifx\ifLONG\undefined#2\else#1\fi}

\newcommand{\documentfinish}{
\ifx\ifLONG\undefined
\bibliographystyle{bibreview} 
\bibliography{\IDvalue,review}  
\end{document}
\else
\addcontentsline{toc}{section}{References}
\providecommand{\href}[2]{#2}\begingroup\raggedright\begin{thebibliography}{10}

\bibitem{ContributionSummary}
V.~Pestun and M.~Zabzine, eds., {\em Localization techniques in quantum field
  theory}, vol.~xx.
\newblock Journal of Physics A, 2016.
\newblock \href{http://arxiv.org/abs/1608.02952}{{\tt 1608.02952}}.
\newblock \url{https://arxiv.org/src/1608.02952/anc/LocQFT.pdf},
  \url{http://pestun.ihes.fr/pages/LocalizationReview/LocQFT.pdf}.

\bibitem{QZWitten:1992xu}
E.~Witten, ``{Two-dimensional gauge theories revisited},''
  \href{http://dx.doi.org/10.1016/0393-0440(92)90034-X}{{\em J. Geom. Phys.}
  {\bf 9} (1992)  303--368},
\href{http://arxiv.org/abs/hep-th/9204083}{{\tt arXiv:hep-th/9204083
  [hep-th]}}.
\bibitem{QZBeasley:2005vf}
C.~Beasley and E.~Witten, ``{Non-Abelian localization for Chern-Simons
  theory},'' {\em J. Diff. Geom.} {\bf 70} (2005) no.~2, 183--323,
\href{http://arxiv.org/abs/hep-th/0503126}{{\tt arXiv:hep-th/0503126
  [hep-th]}}.
\bibitem{QZKapustin:2009kz}
A.~Kapustin, B.~Willett, and I.~Yaakov, ``{Exact Results for Wilson Loops in
  Superconformal Chern-Simons Theories with Matter},''
  \href{http://dx.doi.org/10.1007/JHEP03(2010)089}{{\em JHEP} {\bf 03} (2010)
  089},
\href{http://arxiv.org/abs/0909.4559}{{\tt arXiv:0909.4559 [hep-th]}}.
\bibitem{QZPestun:2007rz}
V.~Pestun, ``{Localization of gauge theory on a four-sphere and supersymmetric
  Wilson loops},'' \href{http://dx.doi.org/10.1007/s00220-012-1485-0}{{\em
  Commun.Math.Phys.} {\bf 313} (2012)  71--129},
\href{http://arxiv.org/abs/0712.2824}{{\tt arXiv:0712.2824 [hep-th]}}.
\bibitem{QZKallen:2011ny}
J.~Kallen, ``{Cohomological localization of Chern-Simons theory},''
  \href{http://dx.doi.org/10.1007/JHEP08(2011)008}{{\em JHEP} {\bf 08} (2011)
  008},
\href{http://arxiv.org/abs/1104.5353}{{\tt arXiv:1104.5353 [hep-th]}}.
\bibitem{QZKallen:2012cs}
J.~K{\"a}ll{\'e}n and M.~Zabzine, ``{Twisted supersymmetric 5D Yang-Mills
  theory and contact geometry},''
  \href{http://dx.doi.org/10.1007/JHEP05(2012)125}{{\em JHEP} {\bf 1205} (2012)
   125},
\href{http://arxiv.org/abs/1202.1956}{{\tt arXiv:1202.1956 [hep-th]}}.
\bibitem{QZDuistermaat1982}
J.~Duistermaat and G.~Heckman, ``On the Variation in the Cohomology of the
  Symplectic Form of the Reduced Phase Space.,'' {\em Inventiones mathematicae}
  {\bf 69} (1982)  259--268. \url{http://eudml.org/doc/142953}.

\bibitem{QZwitten1982}
E.~Witten, ``Supersymmetry and Morse theory,'' {\em Journal of Differential
  Geometry} {\bf 17} (1982) no.~4, 661--692.
  \url{http://projecteuclid.org/euclid.jdg/1214437492}.

\bibitem{ContributionMI}
J.~Minahan, ``Matrix models for 5D super Yang-Mills,'' {\em Journal of Physics
  A} {\bf xx} (2016)  000, \href{http://arxiv.org/abs/1608.02967}{{\tt
  1608.02967}}.

\bibitem{ContributionPA}
S.~Pasquetti, ``Holomorphic blocks and the 5d AGT correspondence,'' {\em
  Journal of Physics A} {\bf xx} (2016)  000,
  \href{http://arxiv.org/abs/1608.02968}{{\tt 1608.02968}}.

\bibitem{QZ1996PhLB..388..753S}
N.~Seiberg, ``{Five-dimensional SUSY field theories, nontrivial fixed points
  and string dynamics},''
  \href{http://dx.doi.org/10.1016/S0370-2693(96)01215-4}{{\em Phys. Lett.} {\bf
  B388} (1996)  753--760},
\href{http://arxiv.org/abs/hep-th/9608111}{{\tt arXiv:hep-th/9608111
  [hep-th]}}.
\bibitem{QZKim:2012gu}
H.-C. Kim, S.-S. Kim, and K.~Lee, ``{5-dim Superconformal Index with Enhanced
  En Global Symmetry},'' \href{http://dx.doi.org/10.1007/JHEP10(2012)142}{{\em
  JHEP} {\bf 1210} (2012)  142},
\href{http://arxiv.org/abs/1206.6781}{{\tt arXiv:1206.6781 [hep-th]}}.
\bibitem{QZHosomichi:2012ek}
K.~Hosomichi, R.-K. Seong, and S.~Terashima, ``{Supersymmetric Gauge Theories
  on the Five-Sphere},''
  \href{http://dx.doi.org/10.1016/j.nuclphysb.2012.08.007}{{\em Nucl.Phys.}
  {\bf B865} (2012)  376--396},
\href{http://arxiv.org/abs/1203.0371}{{\tt arXiv:1203.0371 [hep-th]}}.
\bibitem{QZFestuccia:2011ws}
G.~Festuccia and N.~Seiberg, ``{Rigid Supersymmetric Theories in Curved
  Superspace},'' \href{http://dx.doi.org/10.1007/JHEP06(2011)114}{{\em JHEP}
  {\bf 1106} (2011)  114},
\href{http://arxiv.org/abs/1105.0689}{{\tt arXiv:1105.0689 [hep-th]}}.
\bibitem{QZDumitrescu:2012ha}
T.~T. Dumitrescu, G.~Festuccia, and N.~Seiberg, ``{Exploring Curved
  Superspace},'' \href{http://dx.doi.org/10.1007/JHEP08(2012)141}{{\em JHEP}
  {\bf 1208} (2012)  141},
\href{http://arxiv.org/abs/1205.1115}{{\tt arXiv:1205.1115 [hep-th]}}.
\bibitem{QZDumitrescu:2012at}
T.~T. Dumitrescu and G.~Festuccia, ``{Exploring Curved Superspace (II)},''
  \href{http://dx.doi.org/10.1007/JHEP01(2013)072}{{\em JHEP} {\bf 1301} (2013)
   072},
\href{http://arxiv.org/abs/1209.5408}{{\tt arXiv:1209.5408 [hep-th]}}.
\bibitem{QZClosset:2012ru}
C.~Closset, T.~T. Dumitrescu, G.~Festuccia, and Z.~Komargodski,
  ``{Supersymmetric Field Theories on Three-Manifolds},''
  \href{http://dx.doi.org/10.1007/JHEP05(2013)017}{{\em JHEP} {\bf 1305} (2013)
   017},
\href{http://arxiv.org/abs/1212.3388}{{\tt arXiv:1212.3388 [hep-th]}}.
\bibitem{QZPan:2013uoa}
Y.~Pan, ``{Rigid Supersymmetry on 5-dimensional Riemannian Manifolds and
  Contact Geometry},''
\href{http://arxiv.org/abs/1308.1567}{{\tt arXiv:1308.1567 [hep-th]}}.
\bibitem{QZFriedrichKath}
T.~Friedrich and I.~Kath, ``{Einstein Manifolds of Dimension Five with Small
  First Eigenvalue of the Dirac Operator},'' {\em J. Differential Geometry}
  {\bf 29} (1989)  263--279.

\bibitem{QZQiu:2013pta}
J.~Qiu and M.~Zabzine, ``{5D Super Yang-Mills on $Y^{p,q}$ Sasaki-Einstein
  manifolds},'' \href{http://dx.doi.org/10.1007/s00220-014-2194-7}{{\em Commun.
  Math. Phys.} {\bf 333} (2015) no.~2, 861--904},
\href{http://arxiv.org/abs/1307.3149}{{\tt arXiv:1307.3149 [hep-th]}}.
\bibitem{QZBoyerGalicki}
C.~P. Boyer and K.~Galicki, {\em Sasakian geometry}.
\newblock Oxford Mathematical Monographs. Oxford University Press, Oxford,
  2008.

\bibitem{QZSchmude:2014lfa}
J.~Schmude, ``{Localisation on Sasaki-Einstein manifolds from holomophic
  functions on the cone},''
\href{http://arxiv.org/abs/1401.3266}{{\tt arXiv:1401.3266 [hep-th]}}.
\bibitem{QZAAlekseev}
A.~Alekseev, \href{http://dx.doi.org/10.1007/3-540-46552-9_1}{``Notes on
  Equivariant Localization,''} in {\em Geometry and Quantum Physics},
  H.~Gausterer, L.~Pittner, and H.~Grosse, eds., vol.~543 of {\em Lecture Notes
  in Physics}, pp.~1--24.
\newblock Springer Berlin Heidelberg, 2000.
\newblock \url{http://dx.doi.org/10.1007/3-540-46552-9_1}.

\bibitem{QZKallen:2012va}
J.~K{\"a}ll{\'e}n, J.~Qiu, and M.~Zabzine, ``{The perturbative partition
  function of supersymmetric 5D Yang-Mills theory with matter on the
  five-sphere},'' \href{http://dx.doi.org/10.1007/JHEP08(2012)157}{{\em JHEP}
  {\bf 1208} (2012)  157},
\href{http://arxiv.org/abs/1206.6008}{{\tt arXiv:1206.6008 [hep-th]}}.
\bibitem{QZFigueroaO'Farrill:1999va}
J.~M. Figueroa-O'Farrill, ``{On the supersymmetries of Anti-de Sitter vacua},''
  \href{http://dx.doi.org/10.1088/0264-9381/16/6/330}{{\em Class.Quant.Grav.}
  {\bf 16} (1999)  2043--2055},
\href{http://arxiv.org/abs/hep-th/9902066}{{\tt arXiv:hep-th/9902066
  [hep-th]}}.
\bibitem{QZSalamon}
D.~Salamon, {\em {Spin geometry and Seiberg-Witten invariants}}.
\newblock 1996.

\bibitem{QZHama:2011ea}
N.~Hama, K.~Hosomichi, and S.~Lee, ``{SUSY Gauge Theories on Squashed
  Three-Spheres},'' {\em JHEP} {\bf 1105} (2011)  014,
\href{http://arxiv.org/abs/1102.4716}{{\tt arXiv:1102.4716 [hep-th]}}.
\bibitem{QZImamura:2011wg}
Y.~Imamura and D.~Yokoyama, ``{N=2 supersymmetric theories on squashed
  three-sphere},'' {\em Phys.Rev.} {\bf D85} (2012)  025015,
\href{http://arxiv.org/abs/1109.4734}{{\tt arXiv:1109.4734 [hep-th]}}.
\bibitem{QZImamura:2012xg}
Y.~Imamura, ``{Supersymmetric theories on squashed five-sphere},''
\href{http://arxiv.org/abs/1209.0561}{{\tt arXiv:1209.0561 [hep-th]}}.
\bibitem{QZImamura:2012bm}
Y.~Imamura, ``{Perturbative partition function for squashed $S^5$},''
\href{http://arxiv.org/abs/1210.6308}{{\tt arXiv:1210.6308 [hep-th]}}.
\bibitem{QZWolf:2012gz}
M.~Wolf, ``{Contact Manifolds, Contact Instantons, and Twistor Geometry},''
  \href{http://dx.doi.org/10.1007/JHEP07(2012)074}{{\em JHEP} {\bf 1207} (2012)
   074},
\href{http://arxiv.org/abs/1203.3423}{{\tt arXiv:1203.3423 [hep-th]}}.
\bibitem{QZBaraglia:2014gma}
D.~Baraglia and P.~Hekmati, ``{Moduli Spaces of Contact Instantons},''
\href{http://arxiv.org/abs/1401.5140}{{\tt arXiv:1401.5140 [math.DG]}}.
\bibitem{QZPan:2014nha}
Y.~Pan, ``{Note on a Cohomological Theory of Contact-Instanton and Invariants
  of Contact Structures},''
\href{http://arxiv.org/abs/1401.5733}{{\tt arXiv:1401.5733 [hep-th]}}.
\bibitem{QZQiu:2014cha}
J.~Qiu and M.~Zabzine, ``{On twisted N=2 5D super Yang-Mills theory},''
  \href{http://dx.doi.org/10.1007/s11005-015-0804-8}{{\em Lett. Math. Phys.}
  {\bf 106} (2016) no.~1, 1--27},
\href{http://arxiv.org/abs/1409.1058}{{\tt arXiv:1409.1058 [hep-th]}}.
\bibitem{QZHaydys}
A.~Haydys, ``{Fukaya-Seidel category and gauge theory},'' {\em ArXiv e-prints}
  (2010)  , \href{http://arxiv.org/abs/1010.2353}{{\tt arXiv:1010.2353}}.

\bibitem{QZWitten_FK}
E.~Witten, ``{Fivebranes and Knots},''
\href{http://arxiv.org/abs/1101.3216}{{\tt arXiv:1101.3216 [hep-th]}}.
\bibitem{QZCherkis:2014xua}
S.~A. Cherkis, ``{Octonions, Monopoles, and Knots},''
\href{http://arxiv.org/abs/1403.6836}{{\tt arXiv:1403.6836 [hep-th]}}.
\bibitem{QZKohnRossi}
J.~J. Kohn and H.~Rossi, ``On the Extension of Holomorphic Functions from the
  Boundary of a Complex Manifold,'' {\em Annals of Mathematics} {\bf 81} (1965)
  no.~2, pp. 451--472. \url{http://www.jstor.org/stable/1970624}.

\bibitem{QZEllip_Ope_Cpct_Grp}
M.~F. Atiyah, {\em {Elliptic operators and compact groups}}, vol.~401.
\newblock Springer-Verlag, Berlin, 1974.

\bibitem{QZQiu:2014oqa}
J.~Qiu, L.~Tizzano, J.~Winding, and M.~Zabzine, ``{Gluing Nekrasov partition
  functions},'' \href{http://dx.doi.org/10.1007/s00220-015-2351-7}{{\em Commun.
  Math. Phys.} {\bf 337} (2015) no.~2, 785--816},
\href{http://arxiv.org/abs/1403.2945}{{\tt arXiv:1403.2945 [hep-th]}}.
\bibitem{QZMR2101221}
A.~Narukawa, ``The modular properties and the integral representations of the
  multiple elliptic gamma functions,''
  \href{http://dx.doi.org/10.1016/j.aim.2003.11.009}{{\em Adv. Math.} {\bf 189}
  (2004) no.~2, 247--267}. \url{http://dx.doi.org/10.1016/j.aim.2003.11.009}.

\bibitem{QZTW}
L.~{Tizzano} and J.~{Winding}, ``{Multiple sine, multiple elliptic gamma
  functions and rational cones},'' {\em ArXiv e-prints} (2015)  ,
  \href{http://arxiv.org/abs/1502.05996}{{\tt arXiv:1502.05996}}.

\bibitem{QZPeskinSchroeder}
M.~E. Peskin and D.~V. Schroeder, {\em {An Introduction To Quantum Field
  Theory}}.
\newblock {Frontiers in Physics}. {Westview Press; First Edition}, 1995.

\bibitem{QZ2010arXiv1004.2461S}
J.~{Sparks}, ``{Sasaki-Einstein Manifolds},'' {\em ArXiv e-prints} (2010)  ,
  \href{http://arxiv.org/abs/1004.2461}{{\tt arXiv:1004.2461}}.

\bibitem{QZGauntlett:2004yd}
J.~P. Gauntlett, D.~Martelli, J.~Sparks, and D.~Waldram, ``{Sasaki-Einstein
  metrics on $S^2 \times S^3$},''
  \href{http://dx.doi.org/10.4310/ATMP.2004.v8.n4.a3}{{\em
  Adv.Theor.Math.Phys.} {\bf 8} (2004)  711--734},
\href{http://arxiv.org/abs/hep-th/0403002}{{\tt arXiv:hep-th/0403002
  [hep-th]}}.
\bibitem{QZCvetic:2005ft}
M.~Cvetic, H.~Lu, D.~N. Page, and C.~Pope, ``{New Einstein-Sasaki spaces in
  five and higher dimensions},''
  \href{http://dx.doi.org/10.1103/PhysRevLett.95.071101}{{\em Phys.Rev.Lett.}
  {\bf 95} (2005)  071101},
\href{http://arxiv.org/abs/hep-th/0504225}{{\tt arXiv:hep-th/0504225
  [hep-th]}}.
\bibitem{QZ2001math......7201L}
E.~Lerman, ``Contact toric manifolds,'' {\em J. Symplectic Geom.} {\bf 1}
  (2002) no.~4, 659--828, \href{http://arxiv.org/abs/math/0107201}{{\tt
  arXiv:math/0107201 [math]}}.
  \url{http://projecteuclid.org/getRecord?id=euclid.jsg/1092749569}.

\bibitem{QZMartelli:2005tp}
D.~Martelli, J.~Sparks, and S.-T. Yau, ``{The Geometric dual of a-maximisation
  for Toric Sasaki-Einstein manifolds},''
  \href{http://dx.doi.org/10.1007/s00220-006-0087-0}{{\em Commun.Math.Phys.}
  {\bf 268} (2006)  39--65},
\href{http://arxiv.org/abs/hep-th/0503183}{{\tt arXiv:hep-th/0503183
  [hep-th]}}.
\end{thebibliography}\endgroup

\fi
}

\newcommand{\documentfinishBBL}{
\addcontentsline{toc}{section}{References}
\ifx\ifLONG\undefined
\input{\IDvalue.separate.bbl}
\end{document}
\else
\input{\DIRvalue/\IDvalue.volume.bbl}
\fi
}

\ifx\ifLONG\undefined
\def\volcite#1{Contribution \cite{Contribution#1}}
\else 
\def\volcite#1{Chapter \ref{Chapter#1}}
\fi

\DeclareRobustCommand\QZcal{\mathcal}

\usepackage{hyperref}

\usepackage{fullpage,epsfig,graphics,amsbsy,amssymb,cancel,bbm}
\usepackage{slashed}
\usepackage{graphicx}
\usepackage{amsfonts,mathdots,dsfont}
\usepackage{amssymb,amsmath,amscd,mathrsfs}

\hypersetup{hypertexnames=false}

\usepackage{tikz}

\usetikzlibrary{arrows,chains,matrix,positioning,scopes,snakes}

\makeatletter
\tikzset{join/.code=\tikzset{after node path={%
\ifx\tikzchainprevious\pgfutil@empty\else(\tikzchainprevious)%
edge[every join]#1(\tikzchaincurrent)\fi}}}
\makeatother

\tikzset{>=stealth',every on chain/.append style={join},
         every join/.style={->}}
\tikzset{
    >=stealth',
    punkt/.style={
           rectangle,
           rounded corners,
           draw=black, very thick,
           text width=6.5em,
           minimum height=2em,
           text centered},
    pil/.style={
           ->,
           thick,
           shorten <=2pt,
           shorten >=2pt,}
}

\DeclareMathAlphabet{\mathpzc}{OT1}{pzc}{m}{it}

\newenvironment{QZproof}[1][Proof]{\begin{trivlist}
\item[\hskip \labelsep {\bfseries #1}]}{\end{trivlist}}

\newenvironment{QZexample}[1][Example]{\begin{trivlist}
\item[\hskip \labelsep {\bfseries #1}]}{\end{trivlist}}
\newenvironment{QZremark}[1][Remark]{\begin{trivlist}
\item[\hskip \labelsep {\bfseries #1}]}{\end{trivlist}}

\newcommand{\QZqed}{\nobreak \ifvmode \relax \else
      \ifdim\lastskip<1.5em \hskip-\lastskip
      \hskip1.5em plus0em minus0.5em \fi \nobreak
      \vrule height0.5em width0.5em depth0.00em\fi}
\numberwithin{equation}{section}

\newcommand{\im}{\textrm{Im}\,}
\newcommand{\re}{\textrm{Re}\,}
\newcommand{\ep}{\epsilon}

\begin{document}
\thispagestyle{empty}
\documentheader
\else\chapterheader \fi

\newcommand{\QZBS}{\boldsymbol}
\newcommand{\QZBB}{\mathbb}
\newcommand{\SF}{\mathsf}
\newcommand{\FR}{\mathfrak}
\newcommand{\QZSL}{\textsl}

%%%%%%%%%%% miscellaneous
\newcommand{\QZbea}{\begin{eqnarray}}
\newcommand{\QZeea}{\end{eqnarray}}
\newcommand{\QZbe}{\begin{eqnarray}}
\newcommand{\QZee}{\end{eqnarray}}
\newcommand{\nn}{\notag}
\newcommand{\QZTr}{\textrm{Tr}}
\newcommand{\sTr}{\textrm{sTr}}
\newcommand{\smalint}{{\Large\textrm{$\smallint$}}}
\newcommand{\sbullet}{\textrm{\tiny{\textbullet}}}

\newcommand{\QZbra}{\langle}
\newcommand{\QZket}{\rangle}
\newcommand{\bz}{\bar{z}}
\newcommand{\To}{\Rightarrow}

\newcommand{\reeb}{\textrm{\scriptsize{$R$}}}
\newcommand{\sreeb}{\textrm{\tiny{$R$}}}
\newcommand{\gYM}{g_{\textrm{\tiny{$YM$}}}}

\newcommand{\spinc}{spin${}^c~$}

\section{Introduction}\label{QZsec_intro}
The localisation technique in computing exact partition functions has had a long history by now: from Witten's work on 2D Yang-Mills \cite{QZWitten:1992xu}, then on the 3d side, the non-abelian localisation for Chern-Simons theory \cite{QZBeasley:2005vf}, and a more direct approach by embedding Chern-Simons in the $N=2$ supersymmetric Chern-Simons theory \cite{QZKapustin:2009kz}, and finally on 4d there is the groundbreaking work of Pestun \cite{QZPestun:2007rz}. The techniques used in these calculations simplify as one gains more and more insight into what is essential for the localisation and what are mere frills. For example the work of K\"all\'en \cite{QZKallen:2011ny}, by using a nifty field redefinition, a large part of the work in the calculation of \cite{QZKapustin:2009kz} can be circumvented. The similar field redefinitions were
 later used in the work \cite{QZKallen:2012cs} that started a series of work on the localisation in 5 dimension.
Though the context of localisation may be very different, there is a common thread that unifies all of the above approach, namely the Duistermaat-Heckman formula \cite{QZDuistermaat1982} in equivariant cohomology. Let $(X^{2n},\omega)$ be a $2n$ dimensional closed symplectic manifold, if there is a $U(1)$ action on $X$  with moment map $\mu$. Assuming that this action has only isolated fixed points, then the integral
\begin{eqnarray}
 \int\limits_X\frac{\omega^n}{n!}e^{-\mu}=\sum_i\frac{e^{-\mu(p_i)}}{e(p_i)}\nn
 \end{eqnarray}
can be written as a sum of contributions from the fixed points $\{p_i\}$, and $e(p_i)=\prod\limits_a m_a (p_i)$ is the product of the weights $m_a(p_i)$ of the $U(1)$ action on the tangent space at $p_i$. This formula has a generalisation to the case of just having a vector field $V$ on $X$ with only isolated fixed points \cite{QZwitten1982}, as the whole localisation revolves around this formula, we shall review it quickly here to make the paper self-contained. Let a $V$ be a vector field on $X$ with isolated zeros and assume that it is Killing with respect to a given metric $g$. With this data one can define an operator
\begin{eqnarray}
d_V=d+\iota_V~,\label{QZequiv_diff}
\end{eqnarray}
such that $d_V^2=L_V$. This operator would be the equivariant differential
had $V$ been induced by a $U(1)$ action. Let $\alpha$ be a differential form that is closed under $d_V$, note that $\alpha$
necessarily contains forms of different degrees. The integral of $\alpha$ is then given by
\begin{eqnarray}
\int\limits_X\alpha=\sum_{p_i}\frac{\pi^n~ \alpha|_{p_i}}{{\det}^{1/2}L_V|_{p_i}}~,\label{QZmaster_loc}
\end{eqnarray}
where the sum is over fixed points of $V$. At each $p_i$ the Lie derivative $L_V$ acts as an automorphism of $T_{p_i}X$ and we can compute its determinant.
 The normalized infinitesimal form of $V$ at point $p_i$ is
 \begin{eqnarray}
 V= 2\pi \sum_a m_a \left ( x_a \frac{\partial}{\partial y_a} - y_a \frac{\partial}{\partial x_a} \right )~,
 \end{eqnarray}
  where the positive integers $m_a (p_i)$ are  the weights  of the $U(1)$ action at $p_i$.
The proof of this formula using the Grassmann variables will be given in section \ref{QZsec_localisation}.

The formula (\ref{QZmaster_loc}) is the basis of our localisation technique. In fact, what one shall do is to find, in a given supersymmetric theory, a particular combination of the supersymmetry generator that behaves just like the operator (\ref{QZequiv_diff}) and then apply (\ref{QZmaster_loc}). In the infinite dimensional (path integral) setting, the vector field $V$ is acting on the space of fields, and usually involves a combination of gauge transformation plus a Lie derivative along some vector field on the manifold where our gauge theory is formulated. Our task   is to review
 the details of this procedure for 5d supersymmetric gauge theories. The review is organized as follows: in section \ref{QZsec_Tbs} we review 5d supersymmetric Yang-Mills theory on flat space and on curved spaces. In section \ref{QZsec_Tccc} we turn the susy algebra of section
 \ref{QZsec_Tbs} into the desired form (\ref{QZequiv_diff}) and discuss its natural deformations. In section
  \ref{QZsec_Aotfp} we find the localization locus which gives us an interesting set of differential equations on 5-manifold.
   In section \ref{QZsec_Gfatd} we perform the localization and express the final perturbative answer as the matrix model
    with generalised triple sine function, see \volcite{MI} for  the study  of these matrix models. 
    We also conjecture the full answer for the partition function, see also \volcite{PA} for further discussion.
   Finally in section \ref{QZsec_Aacwfs} we discuss the relation between the curved space computations and   the 1-loop perturbation computation on a flat space.
In the appendices \ref{QZsec_GS} we collect the necessary material on the geometrical setting of the 5d theory, namely the 5d toric Sasaki-Einstein manifolds.

\section{The basic setup}\label{QZsec_Tbs}
\subsection{5d SYM on flat space}

We discuss briefly the setting on the flat space, since later we will extract from our curved space computation certain quantity such as the $\beta$-function, which
can be compared to the explicit 1-loop computation on flat space.

 We are interested in Euclidean version of $N=1$ supersymmetric Yang-Mills theory on $\mathbb{R}^5$ which can be obtained by
  the reduction of 6d $N=1$ theory on $\mathbb{R}^{5,1}$.
The 5d  supersymmetric Yang-Mills action on flat space is
\begin{eqnarray}
S=\frac1{\gYM^2}\QZTr_f\int\limits_{\QZBB{R}^5} d^5x \Big [\frac12 F^{mn}F_{mn}+i\lambda_I\slashed{D}\lambda^I-(D^m\sigma)(D_m\sigma)-\lambda_I[\sigma,\lambda^I]-\frac12D_{IJ}D^{IJ}\Big ] ~,\label{QZaction_flat}
\end{eqnarray}
where $\QZTr_f$ is normalised as $\QZTr_f[t^at^b]=\delta^{ab}/2$ and $D_m$ is covariant derivative.
The various fields are: $F$ is the field strength of the gauge connection $F_{ij}=\partial_{[i}A_{j]}-iA_{[i}A_{j]},~~i,j=1,\cdots,5$;
and $\sigma$ is an adjoint scalar (its kinetic term has the wrong sign because $\sigma$ is in fact the temporal component of the gauge field $\sigma\sim A^0$, since the theory comes from a compactification of a 6d theory on $\QZBB{R}^{1,5}$); while the field $D_{IJ}$ is an auxiliary field in adjoint  that is an isotriplet, i.e. $D_{IJ}=D_{JI}$ and $I,J=1,2$ are the isospin indices.
The fermions $\lambda^I$ above are the gaugini in adjoint, here the pairing of the spinors above uses only transposition
\begin{eqnarray}
 \psi\Gamma^{i_1}\cdots \Gamma^{i_n}\chi\stackrel{\mathrm{def}}{=}\psi^TC\Gamma^{i_1}\cdots \Gamma^{i_n}\chi~,\label{QZspin_pairing_short}
 \end{eqnarray}
where $C$ is the charge conjugation matrix satisfying $C\Gamma^iC^{-1}=(\Gamma^i)^T$ ($C$ is the product of all the gamma matrices that are imaginary or real).
The  $\lambda_I$ are symplectic Majorana spinors satisfying
\begin{eqnarray}
(\lambda^I)^*=\ep_{IJ}C\lambda^J~.\label{QZMajarona}
\end{eqnarray}
Sometimes there is debate about the treatment of the reality condition for fermions when one passes from Lorentz to Euclidean signatures. However
 in the action above, the pairing of spinors uses only the transposition. Since the conjugation of a fermion never appears, while the integration over fermions is
  a formal integral\footnote{By this we mean there is no need to choose a cycle for the integration, in contrast to when one integrates a holomorphic form.}, the problem of how
  to treat the reality condition properly will not affect our calculation.

The field content of 5d SYM can also be understood from the 4d point of view. The 5d N=1 susy reduces to the 4d N=2 susy and the field content is quite familiar. The vector multiplet part of the action written in terms of the 4d N=1 super fields is
\begin{eqnarray}
S=\frac1{4\pi}\im\Big(\frac12\int d^2\theta d^5x~\frac{\partial^2{\QZcal F}}{\partial{\QZcal A}^i\partial{\QZcal A}^j}W^iW^j+\int d^4\theta d^5x~\bar{\QZcal A}^i\frac{\partial{\QZcal F}}{\partial{\QZcal A}^i}\Big)~.\label{QZaction_sf}
\end{eqnarray}
Here $W$ is a spinor fermionic chiral superfield, its leading component is the gaugino and it also contains the self-dual part of the field strength. The field ${\QZcal A}$ is a
chiral superfield that contains the other gaugino and the scalar in the vector multiplet. The object ${\QZcal F}({\QZcal A})$ is called the prepotential and is holomorphic in ${\QZcal A}$.
In 5d the real part of the complex scalar becomes the $5^{th}$ component of the gauge field leaving behind a real scalar that was called $\sigma$ in (\ref{QZaction_flat}).

It is a remarkable feature that in 5d, the prepotential ${\QZcal F}$ has the following most general form \cite{QZ1996PhLB..388..753S}
\begin{eqnarray}
 {\QZcal F}=h_{ij}{\QZcal A}^i{\QZcal A}^j+c_{ijk}{\QZcal A}^i{\QZcal A}^j{\QZcal A}^k~,\label{QZprepostential}
 \end{eqnarray}
where $i$ is the index of the adjoint representation.
For example, in the standard case
\begin{eqnarray} h_{ij}=(\frac{\theta}{2\pi}+\frac{4\pi i}{\gYM^2})\delta_{ij}~.\nn\end{eqnarray}
The cubic term would then contribute a Chern-Simons term \cite{QZKim:2012gu} (remember that the leading component of ${\QZcal A}$ contains the fifth component of the gauge field)
\begin{eqnarray} \frac{c}{6}\QZTr{\QZcal A}^3\to CS_5+\frac{c}{2\pi^2}\QZTr\int\sigma\big( F\wedge*F+(D\sigma)\wedge *(D\sigma)+\cdots\big)~,\label{QZA^3_term}\\
CS_5=\frac{-ic}{24\pi^2}\QZTr \int\limits_{M^5}~\Big (A(dA)^2-\frac{3i}{2}A^3dA-\frac35A^5\Big )~.\nn
\end{eqnarray}
The coupling of the Chern-Simons term, which is proportional $c$, must be quantized. This is one way of seeing that one cannot have any higher power terms in (\ref{QZprepostential}), since those would lead to a Chern-Simons term with a field dependent coupling, which is not allowed.

Note that in flat space, even if one sets the cubic term in (\ref{QZprepostential}) to zero to start with, it will be generated at 1-loop. By dimension counting, $h$ has dimension of mass while $c$ is a number, so $c$ cannot depend on $h$ and hence this is a 1-loop effect only. To perform the actual calculation, one can use the background field method and then we can compare it to our localization result at the flat space limit. The two results agree, but in an indirect way.

\subsection{SYM on the simply connected Sasaki-Einstein manifolds}

The theory (\ref{QZaction_flat}) was constructed  on the five sphere in \cite{QZHosomichi:2012ek}, and a more systematic way of placing a supersymmetric theory on curved space was presented in \cite{QZFestuccia:2011ws}. The general method is that one starts from a suitable supergravity theory and then sends $M_{pl}\to\infty$, i.e. one freezes gravity. It is a large enterprise to study and classify the geometry arising this way that supports at least a fraction of the supersymmetry, see \cite{QZDumitrescu:2012ha,QZDumitrescu:2012at,QZClosset:2012ru} and also \cite{QZPan:2013uoa}.

We will not focus on the most general supersymmetric 5d gauge theories and instead  we shall study the simplest case namely Sasaki-Einstein geometry. It turns out that in 5d one can freeze gravity if one can solve the Killing spinor equation
\begin{eqnarray}
D_m\xi=\pm\frac{i}{2}\Gamma_m\xi_J~,\label{QZkilling_eqn}
\end{eqnarray}
where $D$ is the spin covariant derivative. The number of independent solutions will determine the number of supersymmetry possessed by the theory.
The Killing equation actually leads to Sasaki-Einstein geometry, see \cite{QZFriedrichKath} and also the review in \cite{QZQiu:2013pta}. For example to see that
 it is Einstein, apply (\ref{QZkilling_eqn}) twice, one gets
\begin{eqnarray}
 D_mD_n\xi=-\frac14\Gamma_n\Gamma_m\xi~,\nn
 \end{eqnarray}
whose antisymmetric part forces $R_{mnpq}\Gamma^{pq}\xi=2\Gamma_{mn}\xi$. Multiplying both sides by $\Gamma^n$
\begin{eqnarray}
 \Gamma^nR_{mnpq}\Gamma^{pq}\xi=2\Gamma^n\Gamma_{mn}\xi
~\stackrel{\rm Bianchi}{\To}~R_{mq}\Gamma^q\xi=4\Gamma_m\xi\nn~\To~R_{mn}=4g_{mn}\nn~.\end{eqnarray}
The basic trick of the trade is to construct some tensors out of the Killing spinors, and apply (\ref{QZkilling_eqn}) to determine what sort of differential identities these tensors obey. Then with some luck, one can classify the geometry.

From now on we focus on the simply connected 5d Sasaki-Einstein (SE) manifolds\footnote{By theorem 7.5.27 in \cite{QZBoyerGalicki}, such manifolds are spin}, then the solution can be organized as doublets
\begin{eqnarray}
 D_m\xi_I=\frac{1}{r} t_I^{~J}\Gamma_m\xi_J~,~~~t_I^{~J}=\frac{i}{2}(\sigma_3)_I^{~J}~,~~~(\xi_I\xi_J)=-\frac12\ep_{IJ}~,\label{QZkilling_eqn_I}
 \end{eqnarray}
where we have inserted $r$ as a dimensionful parameter (the size of manifold, which can be ignored for now); and $\sigma_3=\rm{diag}[1,-1]$.

We pause to highlight some key features of the simply connected 5d SE manifolds, which will be used in the formulation of the susy theory, leaving a more detailed review to the appendix.

Out of the solution to (\ref{QZkilling_eqn_I}), one can construct two tensors (we shall leave out $r$ next)
\begin{eqnarray} \reeb^p=\xi_I\Gamma^p\xi^I,~~~J_m^{~\;n}=-2t^{IJ}\xi_I\Gamma_m^{~\;n}\xi_J,\label{QZReeb_and_J}\end{eqnarray}
these quantities satisfy
\begin{itemize}
  \item $\reeb$ has constant norm 1 and it is a Killing vector field, called the \emph{Reeb} vector field
  \item $J$ is horizontal with respect to  $\reeb$, i.e. $\reeb^m J_m^{~\;n}=0$
  \item $J$ defines a complex structure transverse to $\reeb$, i.e.
  $J$ squares to $-1$ restricted to the plane perpendicular to $\reeb$
  \item $Jg$ is the K\"ahler form transverse to $\reeb$
\end{itemize}
The Sasaki condition implies also the equation
\begin{eqnarray}
 \nabla_mJ^p_{~q}=-v^pg_{mq}+\delta^p_m\kappa_q\nn~.
 \end{eqnarray}
That transverse to $\reeb$, there is a K\"ahler structure is particularly important, both in bulding a convenient spin representation and in the index computation
of Schmude \cite{QZSchmude:2014lfa}, see later. We also remark that the transverse K\"ahler structure is K\"ahler-Einstein ($R^{\perp}_{pq}=6g^{\perp}_{pq}$).

\subsection{Field Content and Susy transformation}

In this section we recall the action of 5d SYM for both the vector and hyper-multiplet, and their susy transformation. The same formula is valid for $S^5$ or more general SE manifolds, the only difference is the number of solutions to the Killing spinor equation, i.e. the number of susy. In the former case, one get eight susy while only two for the latter.

The field content of the vector-multiplet is as in the flat space case, the off-shell supersymmetry reads
transformation
\begin{eqnarray}
&&\delta A_m = i\xi_I\Gamma_m\lambda^I~,\nn\\
&&\delta\sigma = i\xi_I\lambda^I~,\nn\\
&& \delta\lambda_I = -\frac12(\Gamma^{mn}\xi_I)F_{mn}+(\Gamma^m\xi_I)D_m\sigma-\xi^JD_{JI}+\frac{2}{r} t_I^{~J}\xi_J\sigma~, \label{QZsusy_vect} \\
&&\delta D_{IJ} = -i\xi_I\Gamma^mD_m\lambda_J+[\sigma,\xi_I\lambda_J]+\frac{i}{r}t_I^{~K}\xi_K\lambda_J+(I\leftrightarrow J)~,\nn
\end{eqnarray}
where $\xi_I$ is a spinor satisfying the Killing equation  (\ref{QZkilling_eqn_I}).
The susy invariant action is
\begin{eqnarray}
&& S_{vec}= \frac{1}{\gYM^2} \int\limits_M\textrm{Vol}_M\,\QZTr\Big[\frac{1}{2}F_{mn} F^{mn} -D_m\sigma  D^m\sigma-\frac12 D_{IJ}D^{IJ}+\frac{2}{r} \sigma t^{IJ}D_{IJ}- \frac{10}{{r}^2}
 t^{IJ}t_{IJ}\sigma^2\nn\\
&&\hspace{2cm}+i\lambda_I\Gamma^mD_m\lambda^I-\lambda_I[\sigma,\lambda^I]-\frac{i}{r}t^{IJ}\lambda_I\lambda_J\Big]~,\label{QZaction_vector}
\end{eqnarray}
where one see that compared to (\ref{QZaction_flat}), certain $1/r$ corrections appeared. Upon sending $r \rightarrow \infty$ we recover the flat action (\ref{QZaction_flat}) and flat supersymmetry transformations.
 We point out a technical detail that the vev $D_{IJ}\sim (2/r)t_{IJ}\sigma$ from solving the eom above is not a susy background, but in contrast $D_{IJ}\sim- (2/r) t_{IJ}\sigma$ is. The difference vanishes in the flat space limit.

%???????????????
%We also write down the essential modification of (\ref{QZA^3_term}) on $S^5$
%\begin{eqnarray} \frac{1}{6}{\QZcal A}^3\to \frac{1}{2\pi^2}\QZTr\int \sigma\big[F\wedge*F+(D\sigma)*(D\sigma)+\frac{R}{12}\sigma^2+\frac{1}{r^2}\sigma^2\cdots\Big]+CS_5.\label{QZA^3_S^5}\end{eqnarray}
%%In fact one simply applies a conformal transformation to $\sigma (D\sigma)^2\to
%%\sigma\big(D\sigma^2+R\sigma^2/12+\sigma^2/r^2)$.
%??????????

\vskip1cm

The hyper-multiplet consists of an $SU(2)_R$-doublet of complex scalars $q^A_I,~~I=1,2$ and an $SU(2)_R$-singlet fermion $\psi^A$, with the reality conditions ($A=1,2,\cdots,2N$)
\begin{eqnarray}
 (q^A_I)^*=\Omega_{AB}\epsilon^{IJ}q^B_J~,~~(\psi^A)^*= \Omega_{AB}C\psi^B~,\label{QZreality-cond-spin}
 \end{eqnarray}
where $\Omega_{AB}$ is the invariant tensor of $USp(2N)$
\begin{eqnarray} \Omega=\left|\begin{array}{cc}
                                                          0 & \mathds{1}_N \\
                                                          -\mathds{1}_N & 0
                                                        \end{array}\right|~,\nn\end{eqnarray}
and $C$ is the charge conjugation matrix as before.

The gauge group will be a subgroup of $USp(2N)$, in particular we consider the hyper-multiplet with the representation $\underline{N}\oplus\underline{\bar N}$ of $SU(N)$, which is embedded in $USp(2N)$ in the standard manner
\begin{eqnarray} U\to \left|\begin{array}{cc}
            U & 0 \\
            0 & U^{-T}
          \end{array}\right|~,~~~U\in SU(N)~.\nn\end{eqnarray}

Suppressing the gauge group index, the on-shell supersymmetry transformations are written as:
\begin{eqnarray}
&&\delta q_I=-2i\xi_I\psi~,\nn\\
&&\delta\psi=\Gamma^m\xi_I(D_mq^I)+i\sigma \xi_Iq^I-\frac{3}{r} t^{IJ}\xi_Iq_J~. \label{QZhyper-tran-noaux}
\end{eqnarray}
The off-shell version of this transformations will be discussed later.
The Lagrangian invariant under (\ref{QZhyper-tran-noaux}) is
\begin{eqnarray}
&&L_{hyp}=\epsilon^{IJ}\Omega_{AB}D_mq_I^A  D^m q_J^{B} -\epsilon^{IJ}q_I^A \sigma_{AC} \sigma^C_{~B}  q_J^B +
\frac{15}{2r^2} \epsilon^{IJ}\Omega_{AB}t^2  q_I^A  q_J^B
 \nn\\
&&\hspace{1cm}-2i \Omega_{AB}\psi^A\slashed{D}\psi^B-2\psi^A\sigma_{AB} \psi^B -4\Omega_{AB}\psi^A\lambda_Iq^{IB}-iq_I^AD_{AB}^{IJ}q_J^B~,\label{QZaction-matter-1}
\end{eqnarray}
where $t^2=t^{IJ}t_{IJ}=1/2$ and $\sigma_{AB} = \Omega_{AC} \sigma^C_{~B}$. The covariant derivative $D$ includes both the Levi-Civita connection and the gauge connection.

Here as in the vector case, the action is written without complex conjugation, and the fermion integrals are done formally.

\section{The cohomological complex}\label{QZsec_Tccc}
\subsection{Finite dimensional toy model}\label{QZsec_localisation}

Here we give a simple proof of (\ref{QZmaster_loc}) in a way that makes its connection to supersymmetry apparent.
We recommend a nice nice review \cite{QZAAlekseev} that covers large part of this section.

Recall the setting of (\ref{QZmaster_loc}) from section \ref{QZsec_intro}. We have a vector field $V$ acting on a manifold $X$. We denote the coordinates of $X$ as $x^i,~i=1, \cdots,  2n$. One can use the fermionic variables $\psi^i$ to represent the 1-forms $dx^i$, and hence a function ${\QZcal O}(x,\psi)$ is just a differential form on $X$. An integral of a differential form is then written as a Grassmann integral
\begin{eqnarray}
 I=\int~ d^{2n} x~d^{2n}\psi ~{\QZcal O}(x,\psi)~.\nn
 \end{eqnarray}
Assume that the differential form $O(x,\psi)$ on $X$ is invariant under an odd symmetry
\begin{eqnarray}
 \delta_V x^i=\psi^i~,~~~~~\delta_V\psi^i=L_Vx^i=V^i(x)~.\label{QZabove_form}
\end{eqnarray}
In fact this complex is nothing but the Cartan formula for the Lie-derivative $L_V=\{d,\iota_V\}$, where $\iota_V:~\Omega^i\to\Omega^{i-1}$ is
the contraction of forms with $V$, represented as $V^i\partial_{\psi^i}$ now.

Pick an odd function $W$  satisfying $\delta_V^2W=0$ then we can insert into the integral a factor
\begin{eqnarray}
I(t)=\int~ d^{2n}x~d^{2n}\psi~{\QZcal O}(x,\psi)e^{-t\delta_V W}~,~~~~\delta_V^2W=0~,\label{QZint_loc_def}
\end{eqnarray}
without changing the value of the integral. The last statement can be seen by differentiating with respect to $t$
\begin{eqnarray}
 \frac{d}{dt}I(t)=-\int~ d^nx~d^n\psi~{\QZcal O}(x,\psi)(\delta_V W)e^{-t\delta_V W}=-\int~ d^nx~d^n\psi~\delta_V\big({\QZcal O}(x,\psi)We^{-t\delta_V W}\big)~,\nn
 \end{eqnarray}
in the last integral, we can replace $\delta_V$ with $d=\psi^i\partial_i$ since the integral will only pick up terms top degree in $\psi$. In this way, one can use Stokes theorem and the result $dI/dt=0$
 follows.

We assume that $\delta_V W$ is a well-behaved function, i.e. providing sufficient damping at infinity and whose critical points are isolated, we can then send $t\to\infty$ in (\ref{QZint_loc_def}) and the integral will be concentrated at the critical points of the even part of $\delta W$
\begin{eqnarray}
\lim_{t\to \infty}\int~ d^{2n}x~d^{2n}\psi~{\QZcal O}(x,\psi)e^{-t\delta_V W}=\sum_{\textrm{cr pt}}\;\frac{\pi^{n}{\QZcal O}_0}{\sqrt{\textrm{det}(\delta_V W)^{''}}}~,\label{QZ-detcom}
\end{eqnarray}
Again so long as $\delta_VW$ is well-behaved in the above sense, the determinant appearing  in (\ref{QZ-detcom}) is (up to a phase) independent of $W$.
 To see this, pick a critical point, say $x=0$ and assume that $\delta_V W$ has the expansion
\begin{eqnarray}
 \delta_V W=c+\frac12g_{ij}x^ix^j+\frac12b_{ij}\psi^i\psi^j+\cdots~,\nn
 \end{eqnarray}
then $\delta_V^2W=0$ implies
\begin{eqnarray} 0=\delta_V^2W=g_{ij}x^i\psi^j+b_{ij}V^i\psi^j+\cdots~,\nn\end{eqnarray}
for this to be zero, one must have $V^i\sim x^j\partial_jV^i+\cdots$, and
\begin{eqnarray}
 g_{ij}=-\partial_i V^kb_{kj}~.\nn
 \end{eqnarray}
This leads to
\begin{eqnarray}
\det g=(-1)^{n}\det V'\det b~,~~~\frac{\textrm{pf}\,b}{\sqrt{\det g}}=\pm\frac{1}{\sqrt{\det dV}}~,\nn
\end{eqnarray}
where $dV$ is the derivative of $V$ at $x=0$, regarded as an endomorphism of $T_0X$.
The precise sign can be worked out, but as it is not crucial to the path integral, we just set it to be $+$. Thus in conclusion
\begin{eqnarray}
\int\limits_X~ {\QZcal O}=\sum_{x_0\in cr\;pt}\frac{\pi^{\frac{\dim X}{2}}{\QZcal O}}{\sqrt{\textrm{det}(dV)}}\Big|_{x_0}~.\label{QZloc_toy_disct}
\end{eqnarray}

There is also holomorphic version of the story. Let  $\delta_V$ and $V$ act holomorphically
\begin{eqnarray}
\delta x^i=\psi^i~,~~\delta x^{\bar i}=\psi^{\bar i}~,~~\delta \psi^i=V^i~,~~\delta \psi^{\bar i}=V^{\bar i}~,\nn
\end{eqnarray}
then we have some modification to the above argument. Assume that $\delta W$ has the expansion
\begin{eqnarray}
 \delta W=c+g_{i\bar j}x^ix^{\bar j}+b_{i\bar j}\psi^i\psi^{\bar j}+\cdots~,\nn\end{eqnarray}
and $\delta^2W=0$ implies
\begin{eqnarray}
0=g_{i\bar j}\psi^ix^{\bar j}+b_{i\bar j}V^i\psi^{\bar j}+c.c+\cdots
=x^i(g_{i\bar j}+b_{j\bar j}(\partial_iV^j))\psi^{\bar j}+\cdots\nn\end{eqnarray}
hence $g_{i\bar j}=-b_{j\bar j}(\partial_iV^j)$,
and the determinant to be computed turns into
\begin{eqnarray}
 \frac{\det b_{j\bar j}}{\det g_{j\bar j}}=(-1)^{\dim_{\QZBB{C}}X}\frac{1}{\det(\partial_iV^j)}=(-1)^{\dim_{\QZBB{C}}X}\frac{1}{\det(\partial_{\bar i}V^{\bar j})}~.\nn
 \end{eqnarray}
This case is applicable to the hyper-multiplet.

In fact the setting can be more general, in that $x$ itself can be both even and odd, with $\psi$ being of the opposite parity.

\begin{QZremark}
Based on the fact that the eventual determinant only depends on the vector field $V$ and 
 not on the details of $W$ (provided of course the appropriate
 $W$ exists), one might tend to skip the step of finding $W$. This is viable in a finite dimensional setting above, as the fixed points are really points.
But when we come to infinite dimensional path integral setting, the fixed points of $V$ are described by certain differential equations in the fields, and so one would prefer to find a $W$ such that its critical points \emph{imply} the given  differential equations, and desirably a bit more, so that one can study effectively the fixed points.
\end{QZremark}

It is straightforward to generalise the above to the case when the zero of $V$ is a submanifold $Z$ of codimension $p$,
\begin{eqnarray}
(\ref{QZloc_toy_disct})=\int\limits_Z\,\frac{\pi^{\textrm{codim}\,Z/2}{\QZcal O}}{\sqrt{\textrm{det}(dV)}}~,\label{QZloc_toy_non_disct}
\end{eqnarray}
where $dV$ is now regarded as an endomorphism of the normal bundle $N_Z$ of $Z$ in $X$.

\subsection{Change of variable}

Our goal next will be to put the vector and hyper multiplet into the complex of the form (\ref{QZabove_form}). We suggest the reader to take a look at the appendix where various geometrical objects of the SE manifold are explained. The most important one we shall use is
the projector
\begin{eqnarray}
P_{\pm}=\frac12(\iota_{\sreeb}\kappa\pm \iota_{\sreeb}*):~~\Omega^2\to \Omega^{2\pm}_H~,\label{QZduality_proj}
\end{eqnarray}
where $\kappa$ is the contact 1-form. In the current setting it is just $\kappa=g\reeb$ and it satisfies
\begin{eqnarray}
\iota_{\sreeb}\kappa=1~,~~\iota_{\sreeb}dk=0~,~~\frac18 \kappa d\kappa d\kappa=\textrm{Vol}~.\nn
\end{eqnarray}
We see that $d\kappa$ is nondegenerate on the plane transverse to $\reeb$ (in fact it is propositional to the transverse K\"ahler form). Further $\Omega^{2\pm}_H$ are the horizontal (anti)-self-dual 2-forms, so the projector (\ref{QZduality_proj}) is the 5d lift of the usual 4d self-duality projector, we will see shortly the 5d instantons are also the lift of the usual 4d anti-self-dual instantons.

Schematically the anti-commutator of two susy transformations is
\begin{eqnarray}
\{\delta_1,\delta_2\}=\textrm{translation}+\textrm{R-rotation}+\textrm{dilatation}+\textrm{gauge transformation}~,\label{QZraw_form}
\end{eqnarray}
where by translation we mean the infinitesimal diffeomorphism along a vector field.
The Killing spinors $\xi$ single out one particular susy $\delta_{\xi}$ that has a simpler anti-commutator
\begin{eqnarray}
\delta_{\xi}^2=\textrm{translation}+\textrm{gauge transformation}~,\label{QZpromised_form}\end{eqnarray}
next we exhibit this for the vector and hyper-multiplet cases.

\smallskip

\noindent\emph {Vector-multiplet}

We use the Killing spinor to turn the gaugino into some odd differential forms. Define
\begin{eqnarray}
 \Psi_m=\xi_I\Gamma_m\lambda^I~,~~~~~\chi_{mn}=\xi_I\Gamma_{mn}\lambda^I+\reeb_{[m}   \xi_I\Gamma_{n]}\lambda^I~,\label{QZfield_redef}
 \end{eqnarray}
the 2-form $\chi$ satisfies the same conditions as $J$:
\begin{eqnarray} \iota_{\sreeb}\chi=0~,~~~~\iota_{\sreeb} * \chi=\chi~.\label{QZhorizontal_self_dual}\end{eqnarray}
This change of variable is invertible
\begin{eqnarray}
\lambda_I=-\frac12\xi^J(\xi_J\Gamma^{mn}\xi_I)\chi_{mn}+(\Gamma^m\xi_I)\Psi_m~,\label{QZredefine_spinor_fine}
\end{eqnarray}
with $\Psi_m$ having 5 components and $\chi_{mn}$ having 3 components.

With the new variables the susy transformation reads (where we just write $\delta$ for $\delta_{\xi}$)
\begin{eqnarray}
&&\begin{array}{ll}
  \delta A = i\Psi~, & \delta \Psi = -\iota_{\sreeb}F+D\sigma~, \\
  \delta \chi = H~, & \delta H=-iL^A_{\sreeb}\chi-[\sigma,\chi]~, \\
  \delta \sigma =-i \iota_{\sreeb}\Psi~, &
\end{array}\nn\\
&&\delta^2=-iL_{\sreeb}+iG_{\Phi}~,~~~\Phi=\sigma+\iota_{\sreeb}A~.\label{QZsusy_vect_twist}\end{eqnarray}
Here $H$ is the bosonic partner of $\chi$ and hence has exactly the same property (\ref{QZhorizontal_self_dual}), explicitly it is related to the auxiliary $D_{IJ}$ as
\begin{eqnarray}
H_{mn}=2(F_H^+)_{mn}+(\xi^I\Gamma_{mn}\xi^J)(D_{IJ}+\frac{2}{r}t_{IJ}\sigma)~,~~~F_H^+=P_+F~.\label{QZD_to_H}
\end{eqnarray}
Further $G_{\Phi}$ is the gauge transformation with respect to  parameter $\Phi$,
defined as\footnote{Our convention is that the Lie algebra  $u(n)$ is given by  Hermitian matrices, and $D=d-iA=d-iA^at^a$ for a basis $\{t^a\}$ of the Lie algebra. This choice engenders awkward $i$'s everywhere, something that we came to regret.}
\begin{eqnarray}
G_{\ep}A=D\ep~,~~~G_{\ep}\phi=i\ep\phi~,\nn
\end{eqnarray}
with $\phi$ in any representation (e.g., if $\phi$ is in the adjoint then $\ep\phi=[\ep,\phi]$). Finally $L^A_{\sreeb} = L_{\sreeb} + i [~,\iota_{\sreeb} A] $ is the Lie derivative combined with gauge transformation. So we see that the square of the susy transformation has the promised form (\ref{QZpromised_form}).

\begin{QZremark}
Intuitively, we can understand the simplification from
(\ref{QZraw_form}) to (\ref{QZpromised_form}) as follows. In our redefinition of fields, we combined the $SU(2)$ doublet $\lambda^I$ with another doublet $\xi_I$, forming a singlet, and so the R-rotation vanishes from
right hand side of (\ref{QZpromised_form}). As for the dilatation, it would have the divergence $\mathrm{div}\, \reeb$ as its parameter, but since $\reeb$ is Killing, this vanishes too.
\end{QZremark}
%\begin{QZremark}\label{QZrmk_susy_bkgd} {\color{red} I am thinking about removing this remark. It is confusing in some way.}
%Looking at the action (\ref{QZaction_vector}), the expectation value of $D_{IJ}$ would be $D_{IJ}\sim 2r^{-1}t_{IJ}\sigma$. But this background is a \emph{not a supersymmetric one} (at least not when the gauge field is horizontal anti-selfdual, which is the ultimate background we will treat in the review). This can be see from the association (\ref{QZD_to_H}), and that $H$ is the susy variation of the fermion $\chi$.
%\end{QZremark}

\smallskip

\noindent \emph{Hyper-multiplet}

Knowing that one should form $SU(2)$ singlets to eliminate R-rotation from the square of the susy transformation, we combine the scalar $q_I$ with the Killing spinors, and leave the fermion $\psi$ alone as it is already a singlet. Thus the twisted hyper-complex is formulated in terms of spinors.
The change of variables reads
\begin{eqnarray} q=\xi_Iq^I~,~~~~~q_I=-2\xi_I q~,\nn \end{eqnarray}
where $q$ is a spinor and we remind the reader about the spinor pairing $\xi_I q\stackrel{def}{=}\xi_I^TCq$.
To see that the above change of variables is invertible one has to use the Fierz identities, see \cite{QZKallen:2012va}.

From the reality condition satisfied by $\xi_I$ and $q_I$ one can see that the spinor field $q$ now satisfies the same reality condition as $\psi$
\begin{eqnarray} (q^A)^*=\Omega_{AB}Cq^B\label{QZreality_q}.\nn\end{eqnarray}
Assuming that the gauge group is $SU(N)\subset USp(2N)$, one can solve this constraint by splitting
\begin{eqnarray}
 q^A\To\left|\begin{array}{c}
           q^{\alpha} \\
           -Cq^*_{\beta}
         \end{array}\right|~,\label{QZrewrite_q}\end{eqnarray}
where $q^\alpha$ is now an unconstrained Dirac spinor transforming in a representation of $SU(N)$, indexed by $\alpha$. The field $\psi$ can be dealt with in exactly the same way.

% The fermion can be written in a similar manner
%\begin{eqnarray} \psi^A=\frac{1}2\left|\begin{array}{c}
%                    \psi^\alpha \\
%                    -C\psi^*_\beta
%                  \end{array}\right|~,\label{QZrewrite_fermion}\end{eqnarray}
%where $\psi^\alpha$ is now an unconstrained Dirac spinor transforming in $\underline N$ (here $\alpha$ is the index for the representation).
%  Analogously we can discuss the adjoint representation of
% $SU(N)$ when two copies of the adjoint are embedded into that of $Sp(N)$.

We will also split $\psi$ according to its chirality under
\begin{eqnarray}
 \gamma_5=-\reeb\cdotp\Gamma~,\label{QZchiralilty}
 \end{eqnarray}
i.e. $\psi=\psi_++\psi_-$, $\gamma_5\psi_{\pm}=\pm \psi_{\pm}$.
Note that the spinor $q$ has $\gamma_5q=+q$
always due to the special property of the Killing spinors (see the review in section 2.3 \cite{QZQiu:2013pta}).
After some massive use of the  Fierz identities and introducing an auxiliary field ${\QZcal F}$ with $-1$ $\gamma_5$-eigenvalue, we get the off-shell
 complex
\begin{eqnarray}
&& \delta q = i\psi_+~,\nn\\
&& \delta \psi_+ = (-L_{\sreeb}^s+G_{\Phi})q~,\nn\\
&& \delta \psi_- ={\QZcal F}~,\nn\\
&& \delta {\QZcal F}=(-iL^s_{\sreeb}+iG_{\Phi})\psi_-~,\label{QZsusy_hyper_twist}\\
&& \delta^2=-iL^s_{\sreeb}+iG_{\Phi}~,\label{QZclosure_hyper}
\end{eqnarray}
where $G_{\Phi}$ is the same gauge transformation as in (\ref{QZsusy_vect_twist}).
We have also introduced the spinorial Lie derivative $L_X^s$, defined for Killing vectors $X$, see \cite{QZFigueroaO'Farrill:1999va}\footnote{There is a sign difference in our second term compared to that of \cite{QZFigueroaO'Farrill:1999va}, the reason is the difference in the convention of the Clifford algebra $\{\Gamma_p,\Gamma_q\}=2g_{pq}$ in this paper, while it is $-2g_{pq}$ there.}
\begin{eqnarray}
&& L_X^s=D_X+\frac18\nabla_{[m}X_{n]}\Gamma^{mn}~,\nn\\
&&[L_X^s,L_Y^s]=L^s_{[X,Y]},~~~~[D_m,L_X^s]=0~,\nn\end{eqnarray}
where $\nabla$ is the Levi-Civita connection and $D$ is the spin covariant derivative.
The last property shows that by using Killing vectors one can generate new solutions of the Killing spinor equation.

\noindent \emph{A Convenient Spin Representation}.

For later computation, we would like to choose a convenient spin representation in terms of anti-holomorphic forms. Let
\begin{eqnarray}
W_{can}=\small{\textrm{$\bigoplus$}}\,\Omega_H^{0,\sbullet}(M)~,\label{QZcan_spin_bundle}
\end{eqnarray}
where $\Omega_H^{0,\sbullet}$ consists of horizontal forms anti-holomorphic with respect to $J$.
One then has a representation of the Clifford algebra: let $\psi$ be any section of $W_{can}$ and $\chi$ a 1-form, define the Clifford action
\begin{eqnarray}
\chi\cdotp\psi=\Bigg\{
                             \begin{array}{cc}
                               \sqrt2 \chi\wedge\psi & \chi\in\Omega_H^{0,1}(M) \\
                               \sqrt2 \iota_{g^{-1}\chi}\psi & \chi\in\Omega_H^{1,0}(M) \\
                               (-1)^{\deg+1}\psi & \chi=\kappa \\
                             \end{array}~.\label{QZcan_spin_rep}
                             \end{eqnarray}
In this way, one has a \spinc-structure whose characteristic line bundle (see chapter 5 in \cite{QZSalamon}) is the anti-canonical line bundle associated with the complex structure $J$.

\begin{QZremark}
In this representation, the two Killing spinors are (0,0) and the (0,2) forms respectively. In particular, that a non-vanishing section of (0,2) forms exists follows from the triviality of the canonical bundle of the horizontal complex structure. The latter in turn follows from the K\"ahler Einstein condition: the curvature of the canonical bundle is the (1,1) part of the Ricci tensor which is proportional to $d\kappa$ and hence is trivial.
\end{QZremark}

Thus one has a representation where $q,\psi_+\in \Omega_H^{0,0}\oplus \Omega_H^{0,2}$, and $\psi_-,{\QZcal F}\in \Omega_H^{0,1}$. Furthermore with SE metric the spinorial Lie derivative is related to the usual Lie derivative as
\begin{eqnarray}
L_X^s=L_X+if_X~,\label{QZL_X_L_X^s}
\end{eqnarray}
where $f_X$ is a real constant. In the toric SE case $f_X$ can be read off easily from the toric data and $f_{\sreeb}=3/2$. To summarize, the hyper complex reads
\begin{eqnarray}
&& \delta q = i\psi_+~,~~~\delta \psi_+= (-L_{\sreeb}-if_{\sreeb}+G_{\Phi})q~,\nn\\
&& \delta \psi_- ={\QZcal F}~,~~~\delta {\QZcal F}=i(-L_{\sreeb}-if_{\sreeb}+G_{\Phi})\psi_-~.\label{QZhyper_twist_form}
\end{eqnarray}
We need also the formula for the spinor pairing for later use. If one works through the spinor-form correspondence, the pairing is
\begin{eqnarray} \begin{array}{c|c|c}
\textrm{as spinors} & \xi^TC\eta & \xi^{\dag}\eta \\
\hline
\textrm{as forms} & -(-1)^{d(d-1)/2}\frac{\xi\wedge\eta}{2\bar\varrho} & \xi^*\wedge*\eta \end{array}\label{QZspin_pairing_form}
\end{eqnarray}
where $d=\deg\xi$ and $\bar\varrho$ is the nowhere vanishing section of $\Omega_H^{0,2}$, which exists in the SE geometry, and assume that it is normalised to be of norm 1. In the text, the first paring is denoted as just $\xi\eta$, and we denote it in the form language as $\QZbra\xi,\eta\QZket$, this pairing is known as the Mukai pairing.

\subsection{Deformation of the complex}\label{QZsec_bla_bla}
One of the advantages of reformulating the susy complex in terms of differential forms is that there is natural deformation of the cohomological complex.

Looking at (\ref{QZsusy_vect_twist}), one has the freedom to deform $\reeb$, and in the case of toric SE geometry, the deformation has a very simple parameterisation, see later.
 Putting aside some positivity conditions, the deformation is valid provided one also deforms $\kappa$ and subsequently the horizontal plane correspondingly. One can allow $\reeb$  to have
 a small imaginary part in order to get the stronger locaisation locus. The deformed complex looks exactly the same as (\ref{QZsusy_vect_twist}) so we shall not write it again.

In a series of works \cite{QZHama:2011ea, QZImamura:2011wg, QZImamura:2012xg,QZImamura:2012bm}, one tried to set up susy theories on the squashed three (five) sphere. These manifolds are topologically the same but the metric is no longer SE, so the Killing equations (\ref{QZkilling_eqn}) must be modified. In other words, more background fields from the supergravity multiplet have to be turned on, and these fields modify the
 right hand side of (\ref{QZkilling_eqn}). The net result is that one gets a fairly complicated susy theory, but if one tries to rewrite them in terms of differential forms (\ref{QZsusy_vect_twist}), no change occurred other than replacing $\reeb$ with the deformed one.
 %\footnote{and it has been recently communicated to us by YW. Pan that the same is true for more general 5d geometry supporting susy, namely, all deformations reduces to the deformation of $\reeb$.}.
 Thus as far as computation is concerned, one can take (\ref{QZsusy_vect_twist}) as the starting point.

The deformation of the hyper-multiplet complex is a bit more tricky. We can take the formulation in (\ref{QZsusy_hyper_twist}) and deform $\reeb$ as before, keeping in mind that the chirality operator $\gamma_5=-\reeb\cdotp\Gamma$ has to be deformed accordingly. The only problem is that the spinorial Lie derivative $L_{\sreeb}^s$ depends on the choice of the spin connection, and thus
 the deformation seems less canonical. Alternatively, one can take the reformulation of hypermultiplet in terms of differential forms (\ref{QZhyper_twist_form}) as the starting point  with the Lie derivative acting on forms canonically.  The only remaining problem is to determine the shift $f_{\sreeb}$ and this can be done by using the consistency checks for SE metric.

In particular, we shall deform the metric and horizontal complex structure for the SE manifolds; in the toric SE case, these deformations are easily parameterized, see appendix \ref{QZsec_TSm}. Now we focus on the toric case, by assumption we still have a nowhere vanishing section $\varrho\in\Omega^{0,2}_H$, which shall be constructed also in the appendix, and we show that
\begin{eqnarray} 2if_{\sreeb}\varrho=L_{\sreeb}\varrho~.\label{QZR-charge}\end{eqnarray}
In particular, when the metric is SE, one always has $L_{\sreeb}\varrho=3i\varrho$ and
 so $f_{\sreeb}=3/2$ agreeing with the shift for SE case. And now we use (\ref{QZR-charge}) as a definition
 of the shift $f_{\sreeb}$. Later we will see that this shift has some other virtues.

\section{Analysis of the fixed points}\label{QZsec_Aotfp}
As we did for  the toy model section \ref{QZsec_localisation}, we have to find an appropriate functional $W$ and
 deform the action by $-t\int\delta W$ so as to localize the path integral on the fixed points of the vector field $\delta^2=-iL^{(s)}_{\sreeb}+iG_{\Phi}$.

\subsection{Vector multiplet and contact instantons}
\noindent \emph{An observable}

Beside the supersymetric Yang-Mills action the vector multiplet also possesses an observable that is $\delta$-closed but not $\delta$-exact
\begin{eqnarray}
 &&{\QZcal O}=CS_{3,2}(A+\kappa\sigma)+i\QZTr\int \kappa \wedge d\kappa \wedge\Psi \wedge \Psi~,\label{QZobservable}\\
&&CS_{3,2}=\QZTr\int d\kappa\wedge(A\wedge dA-\frac{2i}{3}A\wedge A\wedge A)=\QZTr\int \kappa\wedge F\wedge F~.\nn\end{eqnarray}
The bosonic part of ${\QZcal O}$ reads
\begin{eqnarray} {\QZcal O}|_{bos}=\QZTr\int \kappa \wedge F\wedge F+2\sigma \kappa \wedge d\kappa \wedge  F+  \sigma^2\kappa \wedge  d\kappa \wedge d\kappa ~.\nn\end{eqnarray}
One can in fact also write an observable associated with the 5d Chern-Simons term, see \cite{QZKallen:2012cs}.

\smallskip

Next we collect the bosonic part of the classical action (\ref{QZaction_vector}) (set $r=1$)
\begin{eqnarray}
&&S_{vec}\big|_{bos}=\QZTr\int \iota_{\sreeb}F\wedge *(\iota_{\sreeb}F)-\kappa \wedge  F\wedge F-(D\sigma)\wedge *(D\sigma) \label{QZSvec_bos}\\
&&\hspace{1cm} -\frac12H\wedge *H+2\kappa \wedge F\wedge H+\frac1r\sigma\kappa \wedge d\kappa \wedge H-\frac{2}{r}\sigma \kappa \wedge d\kappa \wedge
 F-\frac8{r^2}\sigma^2 \kappa \wedge d\kappa \wedge d\kappa ~.\nn \end{eqnarray}
Now we can choose $W$
\begin{eqnarray} W_{vec}(s)=\QZTr\Big[\Psi\wedge *(-\iota_{\sreeb}F-D\sigma)-\frac12\chi\wedge *H+2\chi \wedge *F+ s  \sigma \kappa \wedge d\kappa \wedge \chi\Big]~,\nn
\end{eqnarray}
where $s$ is some parameter and $F_H^{+}=P_+F$, with $P_+$ defined in (\ref{QZduality_proj}). We can check that the classical action is reproduced as
\begin{eqnarray} S_{vec}=-{\QZcal O}+\int\delta W_{vec}(1)~.\nn\end{eqnarray}

But for the deforming part, we take $t\int W_{vec}(0)$, and it is easy to check that
\begin{eqnarray}
\int\delta W_{vec}(0)\big|_{bos}=
\QZTr\int \iota_{\sreeb}F\wedge *(\iota_{\sreeb}F)-(D\sigma)\wedge *(D\sigma)-\frac12H\wedge *H+2F_H^+ \wedge *H~.\label{QZS_vec_def_bos}
\end{eqnarray}
The fields $H$ and $\sigma$ should be to Wick rotated $H\rightarrow iH$ and $\sigma\rightarrow i\sigma$,  in order to have a positive kinetic term and stronger localization locus.
 Next we integrate out $H$ leaving a perfect square. Thus the localisation locus is
\begin{eqnarray}
F_H^+=0~,~~~\iota_{\sreeb} F =0~,~~~D\sigma=0~.\label{QZgeneral-vect-loc1}
\end{eqnarray}
The first two equations came to be called the 'contact instanton' and they can be combined in one equation
\begin{eqnarray}
  *F = - \kappa \wedge F~,\label{QZcontact-instanton}
\end{eqnarray}
while the last equation in (\ref{QZgeneral-vect-loc1}) says $\sigma$ is a covariant constant.
%\begin{QZremark}
%To explain the choice of the parameter $s$, we note that with (\ref{QZS_vec_def_bos}), the v.e.v of $H$ is $H\sim F_H^+=0$, consequently $\delta \chi=0$ and this is a supersymmetric background.
%This is not true for the classical action, compare also with the remark on page \pageref{rmk_susy_bkgd}. {\color{red} maybe move partially this remark here and rewrite it}
%\end{QZremark}

\begin{QZremark}
Recall in the 4d case, (anti)-self-duality of $F$ would imply the Yang-Mills equation. In 5d same thing happens
\begin{eqnarray} D_A(*F)=D_A(\kappa\wedge F)=d\kappa\wedge F-\kappa\wedge D_AF=d\kappa\wedge F~,\nn\end{eqnarray}
but the rhs is zero
\begin{eqnarray} d\kappa\wedge F=(\iota_{\sreeb}*d\kappa)\wedge F
=\iota_{\sreeb}(*d\kappa\wedge F)=\iota_{\sreeb}(d\kappa\wedge *F)=d\kappa\wedge\iota_{\sreeb}F=-d\kappa\wedge F~.\nn\end{eqnarray}
We leave it to the reader to check that the same equation (\ref{QZgeneral-vect-loc1}) but with $F_H^-=0$ will not imply the Yang-Mills equation.
\end{QZremark}
\begin{QZremark}
We comment also that (\ref{QZgeneral-vect-loc1}) is not an elliptic system, so studying its deformation is slightly unconventional (see \cite{QZWolf:2012gz, QZBaraglia:2014gma, QZPan:2014nha}). However, one can embed this set of equations into another set \cite{QZQiu:2014cha}, which is a 5d lift of the Vafa-Witten equation, now called the Haydys-Witten equation \cite{QZHaydys,QZWitten_FK,QZCherkis:2014xua}. The latter set has interesting vanishing theorems so is perhaps better adapted for studying the moduli problem.
\end{QZremark}

The Yang-Mills action saturates a bound at the instanton background
\begin{eqnarray} \int F\wedge *F&=&\int (\kappa\iota_{\sreeb}F+F_H)\wedge*(\kappa\iota_{\sreeb}F+F_H)=\int (\iota_{\sreeb}F)\wedge*(\iota_{\sreeb}F)+F_H^+\wedge*F_H^++F_H^-\wedge*F_H^-\nn\\
&=&\int (\iota_{\sreeb}F)\wedge*(\iota_{\sreeb}F)+2F_H^+\wedge*F_H^+-\kappa \wedge F\wedge F,\label{QZcompare_II}
\end{eqnarray}
where we have used the orthogonality of different subspaces.
The term $\int\kappa\wedge F\wedge F$ provides a weighting for the instantons. Since this term is not topological, so it is not immediately clear that its value is bounded away from zero. This gap is important in that it allows us to take the large $N$ limit and decouple the instanton sector, see \volcite{MI}.  In the simple case of a round $S^5$, it is possible to further analyse the contact instanton configuration, and show that for $SU(2)$ gauge group the contact instantons are in 1-1 correspondence with the instantons on $\QZBB{C}P^2$, see section 3.2.2. of \cite{QZKallen:2012va}\footnote{Due to a historical accident, the choice of volume form in \cite{QZKallen:2012va} is opposite to the current one. The reader should bear this in mind when comparing results between the two papers, especially some anti-self-dualities there will become self-dualities here.}. So in this case, we do have a gap and this gap will be stable against small perturbations of the geometry. For the general case, we believe that if one carries out some careful analysis, one can show the existence of the gap, but we did not investigate it any further.

The round $S^5$ case is special because the Reeb vector field $\reeb$ forms closed orbits everywhere, in fact it is the $U(1)$ rotation along the fibre of the Hopf fibration $U(1) \to S^5\to \QZBB{C}P^2$. However what is more interesting is the opposite extreme: when the Reeb flows are not closed except at a few isolated loci. In this situation, the instanton partition functions are conjectured to concentrate along those few orbits, this conjecture is supported by evidence from the perturbative sector, see \volcite{PA}. For the rest of the review, we focus on the perturbative, i.e. zero instanton sector. To summarise, the localisation locus for the vector multiplet is
\begin{eqnarray}
A=0~,~~\sigma=a=\textrm{const}\in i\FR{g}~,~~\Psi=0~.\label{QZconfig_perturb}
\end{eqnarray}
Evaluating the classical action (\ref{QZSvec_bos}) at this background we get
\begin{eqnarray}
S_{vec} =-\QZTr\int \kappa(d\kappa)^2\sigma^2=-8\textrm{Vol}\,\QZTr[a^2]~,\label{QZGaussian_damping}\end{eqnarray}
this quadratic term will be the Gaussian damping for the matrix model resulting from localisation.

\subsection{Hyper-multiplet and vanishing theorems}\label{QZsec_Hmavt}

As explained at the end of section \ref{QZsec_bla_bla}, if one sticks to the SE geometry, then one can use (\ref{QZsusy_hyper_twist}) as his starting point, while if one deforms the SE geometry then (\ref{QZhyper_twist_form}) is a more convenient starting point.

\smallskip

\noindent \emph{We first deal with the case of SE metric}, i.e. (\ref{QZsusy_hyper_twist}), where one can give a concise proof of a vanishing theorem.
We can add the following exact term $-t\delta \int W_{hyp}$ to the path integral, where
\begin{eqnarray}
 W_{hyp}=\frac12\Omega_{AB}\Big[\psi^A_+\,(-L^{As}_{\sreeb}-G_{\sigma})q^B+\psi^A_-\,{\QZcal F}^B+2i\psi^A_-\slashed{D}q^B\Big]~,\nn
 \end{eqnarray}
where the notation $L_X^{As}$ denotes the spinorial Lie derivative coupled to the gauge potential.
Note that the last two terms are designed to produce a kinetic term for $q$ and the fermions.
The bosonic part of $\delta W_{hyp}$ is
\begin{eqnarray}
 \delta W_{hyp}\big|_{\textrm{bos}}=\frac12\Omega^{AB}\Big[(L^{As}_{\sreeb}q)^A(L^{As}_{\sreeb}q)^B-(G_{\sigma}q)^A
(G_{\sigma}q)^B+{\QZcal F}^A{\QZcal F}^B+2i{\QZcal F}^A\slashed{D}q^B\Big]~.\label{QZalso_applies}
\end{eqnarray}
In the last term, only the negative $\gamma_5$ chirality part of $\slashed{D}q$ will survive since $\gamma_5{\QZcal F}=-{\QZcal F}$. Now integrating out ${\QZcal F}$ produces a good kinetic term $(\slashed{D}q)^2$. To this end, it is now convenient to take the gauge group to be $SU(N)$ and solve the constraint as in (\ref{QZrewrite_q}), then (\ref{QZalso_applies}) turns into
\begin{eqnarray}
\delta W_{hyp}\big|_{bos}=(L^{As}_{\sreeb}q)^{\dag}L^{As}_{\sreeb}q-q^{\dag}\sigma^2q+{\QZcal F}^{\dag}\,{\QZcal F}+i\Big[{\QZcal F}^{\dag}\slashed{D}q+c.c\Big]~,\label{QZhyper_exact_II}
\end{eqnarray}
where we have used the same symbol for the fields before and after the rewriting  (\ref{QZrewrite_q}). Now we can integrate over ${\QZcal F}$ and get
\begin{eqnarray}
W_{hyp}\big|_{bos}=(\slashed{D}-\frac14\slashed{J}q)^{\dag}(\slashed{D}-\frac14\slashed{J}q)-q^{\dag}\sigma^2q~,\nn
\end{eqnarray}
where the key relation used in the step above is
\begin{eqnarray}
 \frac{1}{2}(1+\gamma_5)\big(\slashed{D}-\frac14\slashed{J}\big)q=-L_{\sreeb}^{As}q.~\nn
 \end{eqnarray}

Remembering that $\sigma$ is Wick rotated, we get the localisation locus
\begin{eqnarray} \big(-\frac14\slashed{J}+\slashed{D}\big)q=0~,~~~\sigma q=0.\label{QZlocus_hyper}\end{eqnarray}
We prove next that this set of conditions implies $q=0$.
\begin{QZproof}
We start from the equation $(\slashed{D}-\slashed{J}/4)q=0$ and so
\begin{eqnarray}
 0=(\slashed{D}-\slashed{J}/4)^2q=(\slashed{D}^2-\frac{1}{4}\slashed{J}\slashed{D}-\frac{1}{4}\slashed{D}\slashed{J}+\frac{1}{16}\slashed{J}^2)q=(\slashed{D}^2-\frac{1}{4}\slashed{D}\slashed{J})q~.\nn\end{eqnarray}
Now put this under the integral
\begin{eqnarray}
0=\int q^{\dag}(-\slashed{D}^2+\frac{1}{4}\slashed{D}\slashed{J})q\stackrel{ibp}{=}\int q^{\dag}(-\slashed{D}^2-\frac{1}{4}\overleftarrow{\slashed{D}}\slashed{J})q
=\int q^{\dag}(-\slashed{D}^2+\frac{1}{16}\slashed{J}^2)q~,\nn\end{eqnarray}
note that in our convention the gamma matrices are hermitian, so $\Gamma_p^{\dag}=\Gamma_p,~\Gamma_{pq}^{\dag}=-\Gamma_{pq}$.
The two terms in the integral are
\begin{eqnarray}
\slashed{D}^2=D^2-5-\frac{i}{2}\slashed{F}~,~~~~~\slashed{J}^2=-8(1+\gamma_5)\nn~.
\end{eqnarray}
We also put the gauge field in an instanton configuration (\ref{QZgeneral-vect-loc1}). Then we have
\begin{eqnarray}
q^{\dagger}\slashed{F}q=q^{\dagger}(F_H^+)_{mn}\Gamma^{mn}q=0~,\nn
\end{eqnarray}
since $q^{\dag}\Gamma^{mn}q$ is horizontal self-dual and $F_H^+=0$.
Assembling everything altogether
\begin{eqnarray}
0=\int q^{\dagger}\big(\frac{1}{16}\slashed{J}^2-D^2+5\big)q=\int q^{\dagger}\big(-D^2+4\big)q=\int (D_mq)^{\dagger}(D^mq)+4\int q^{\dagger}q~.\nn\end{eqnarray}
So we must have $q=0$\QZqed
\end{QZproof}
The key perennial trick is to relate two quadratic differential operators, e.g. $\slashed{D}^2$ and $D*D$ hoping to produce some constant terms. This technique will be exploited again shortly.

\smallskip

\noindent \emph{Now we deviate from the SE metric}, but keeping the topology type. It is more convenient to do so in the form formulation of (\ref{QZhyper_twist_form}). Remember that $q,\psi_+\in\Omega_H^{0,0}\oplus\Omega_H^{0,2}$ and $\psi_-,{\QZcal F}\in\Omega^{0,1}_H$, we denote the 0-form component of $q,\psi_+$ as $h,\lambda$ and the 2-form component as $B,\Sigma$, while $\psi_-,{\QZcal F}$ are always 1-forms, we still call them $\psi_-,{\QZcal F}$.

We add an exact term $-t\delta \int W_{hyp}$, which is the same as
(\ref{QZhyper_exact_II}) except the appropriate replacements
\begin{eqnarray}
 \slashed{D}q\To (Dh)^{0,1}-4D^{\dag}B;~~~L_{\sreeb}^{As}\To L_{\sreeb}^A+if_{\sreeb}~.\nn
 \end{eqnarray}
After integrating out ${\QZcal F}$, the bosonic part of $\delta W_{hyp}$ is\footnote{Here $D^{\dag}$ is the adjoint of $D$, and $B^{\dag}$ is the hermitian conjugate of $B$, hopefully, there will be no confusion.}
\begin{eqnarray}
 \delta W_{hyp}\big|_{\textrm{bos}}&=&
\int\big((-L^A_{\sreeb}-if_{\sreeb})q\big)^{\dag}*(-L^A_{\sreeb}-if_{\sreeb})q-( G_{\sigma}q)^{\dag}*G_{\sigma}q\nn\\
&&+(D^{0,1}h-4D^{\dag}B)^{\dag}*(D^{0,1}h-4D^{\dag}B)~.\nn\end{eqnarray}
So the localisation locus, written in differential forms, is
\begin{eqnarray}
 (L^A_{\sreeb}+if_{\sreeb})h=(L^A_{\sreeb}+if_{\sreeb})B=0,~~ D^{0,1}h-4D^{\dag}B=0~.\label{QZlocalisation_hyper_form_I}
 \end{eqnarray}

\smallskip

Next we give a convenient criteria for the vanishing of all fields in the hypermultiplet. Consider the integral
%\begin{eqnarray} 0&=&\int (D^{0,1}h-4D^{\dag}B)^{\dag}*(D^{0,1}h-4D^{\dag}B)+8B^{\dag}(-L^A_{\sreeb}+if_{\sreeb})(L^A_{\sreeb}+if_{\sreeb})B\nn\\
%&=&||D^{0,1}h||^2+16||D^{\dag}B||^2+8(||L^A_{\sreeb}B||^2-||f_{\sreeb}B||^2).\label{QZtemp_I}\end{eqnarray}
\begin{eqnarray} 0=\int (D^{0,1}h-4D^{\dag}B)^{\dag}*(D^{0,1}h-4D^{\dag}B)=||D^{0,1}h||^2+16||D^{\dag}B||^2.\label{QZtemp_I}\end{eqnarray}
Note that the cross term between $D^{0,1}h$ and $D^{\dag}B$ can be shown to vanish by using (\ref{QZgeneral-vect-loc1}).

Now focus on the last three terms, we apply a Weizenbock formula (see (14) in sec.2.5 of \cite{QZQiu:2014cha})
\begin{eqnarray} &&||D^{\dag}B||^2+\frac12||L_{\sreeb}^AB||^2=||DB||^2-\frac12||L_{\sreeb}^AB||^2\nn\\
&&\hspace{3cm}=\int\,iF^a_{~b}*(B^{\dag}_a\times B^b)-\frac14(8-s)B^{\dag}*B+\frac14\QZbra \nabla B^{\dag},\nabla B\QZket,\label{QZWeizenbock_III}\end{eqnarray}
where $s$ is the scalar curvature, $a,b$ are indices in the fundamental of $SU(N)$, $\QZbra-,-\QZket$ is defined in (\ref{QZinner_product}) and the $\times$ operation is defined as
\begin{eqnarray} \Omega_H^{2+}\ni (X\times Y)_{mn}=X_{mp}Y_n^{~p}-X_{np}Y_m^{~\;p},~~~X,Y\in\Omega_H^{2+}.\nn\end{eqnarray}
The term in (\ref{QZWeizenbock_III}) involving the curvature vanishes, since $B^{\dag}_a\times B^b$ is horizontal self-dual and $F$ is horizontal anti-self-dual. And the last term in (\ref{QZWeizenbock_III}) can be broken further into
\begin{eqnarray} \QZbra \nabla B^{\dag},\nabla B\QZket&=&\QZbra (\nabla B^{\dag})_H,(\nabla B)_H\QZket+\QZbra(L^A_{\sreeb}-2i)B^{\dag},(L^A_{\sreeb}+2i)B\QZket+2\QZbra B^{\dag},B\QZket\nn\\
&=&\QZbra (\nabla B^{\dag})_H,(\nabla B)_H\QZket+(2+(2-f_{\sreeb})^2)\QZbra B^{\dag},B\QZket\nn.\end{eqnarray}
Thus (\ref{QZtemp_I}) equals (using (\ref{QZlocalisation_hyper_form_I}))
\begin{eqnarray} \ref{QZtemp_I}=||D^{0,1}h||^2+16\int \big(\frac{s}{4}+1-2f_{\sreeb})B^{\dag}*B+\frac14\QZbra (\nabla B^{\dag})_H,(\nabla B)_H\QZket.\nn\end{eqnarray}
so we can conclude the vanishing of $B$ if $s+4-8f_{\sreeb}>0$ everywhere.

Assuming that the condition above is satisfied and so $B=0$.
To prove the vanishing of $h$, we need a trick, note that $\varrho(L^A_{\sreeb}+if_{\sreeb})h=(L^A_{\sreeb}-if_{\sreeb})(\varrho h)$, then we can combine $\bar\varrho \bar h$ into a 2-form. The equation (\ref{QZlocalisation_hyper_form_I}) plus a choice of $\varrho$ such that $D^{0,1}\varrho=0$ leads to that
\begin{eqnarray} D^{(0,1)}(\varrho h)=0=(L^A_{\sreeb}-if_{\sreeb})(\varrho h)\nn\end{eqnarray}
The rest of the treatment for $h$ will be as for $B$. %, i.e. one applies the Weizenbock formula to
One has
\begin{eqnarray} 0=||D^{0,1}(\varrho h)||^2=||D(\varrho h)||^2-||L_{\sreeb}^A\varrho h||^2.\nn\end{eqnarray}
Now apply the second half of Weizenbock formula (\ref{QZWeizenbock_III}), and arrives at
\begin{eqnarray} 0=\int \big(\frac{s}{4}+1-2f_{\sreeb})(\varrho h)*(\varrho h)^{\dag}+\frac14\QZbra (\nabla \varrho h)_H,(\nabla (\varrho h)^{\dag})_H\QZket\nn\end{eqnarray}
and hence the same vanishing condition.
%\begin{eqnarray} 0=\int \big(\frac14(s-8)-\frac12f_{\sreeb}^2)(\varrho h)^{\dag}*(\varrho h)-|h|^2(D^{1,0}\varrho)^{\dag}*(D^{0,1}h)+\frac14((D\varrho h)^{\dag})^{pqr}(D\varrho h)_{pqr}.\nn\end{eqnarray}
So the conclusion is that if
\begin{eqnarray} s+4-8f_{\sreeb}>0\label{QZvanishing_condition}\end{eqnarray}
then the hyper-multiplet vanishes at the localisation locus.
Again at the SE point, the lhs above equals $20+4-8\cdot3/2=12$ corroborating with the direct proof after (\ref{QZlocus_hyper}).
Now one can perturb the geometry in an open neighbourhood round the SE point and still retain the vanishing result.
We have not carried out the detailed calculation of $s$ away from the SE point, but it is likely that, for the type of deformation we consider in this paper, this condition is true always.
\begin{QZremark}
The sign in front of $if_{\sreeb}$ in (\ref{QZlocalisation_hyper_form_I}) is crucial. In fact, we shall see that the equation
\begin{eqnarray} (L_{\sreeb}-ic)h=D^{0,1}h=0\nn\end{eqnarray}
has non-trivial solutions for infinitely many positive values of $c$. These solutions represent the Kohn-Rossi cohomology. But when $c\geq0$, the vanishing theorem is rendered impotent, since one needs to replace the combination $s+4-8f_{\sreeb}$ in (\ref{QZvanishing_condition}) with
\begin{eqnarray} s+4-8(c+2f_{\sreeb})\nn\end{eqnarray}
which can just escape the vanishing theorem, say, at the SE point for $c\geq0$. This is a nice consistency check on our manipulations with the Weizenbock formula.
\end{QZremark}

\section{Gauge fixing and the determinant}\label{QZsec_Gfatd}

Now we are ready to apply the abstract model in section \ref{QZsec_localisation} to the gauge theory and we follow closely the original work by Pestun \cite{QZPestun:2007rz}.
The hyper complex (\ref{QZsusy_hyper_twist}) is perfectly analogous to (\ref{QZabove_form}), however the vector complex (\ref{QZsusy_vect_twist}) is not, in that the combination $\Phi=\iota_{\sreeb}A+\sigma$ has no susy variation. Besides this, there is the problem of gauge fixing. We first fix the gauge bundle to be topologically trivial, since we are only interested in the zero instanton sector, in particular, the connection $A$ is a global adjoint 1-form.

We will take a shortcut and arrive at the answer faster though admittedly less rigorously. Up to gauge transformation, the fixed points are given by (\ref{QZconfig_perturb}), and we use the gauge freedom to fix $\Phi$ at $\Phi=a$. Doing this will incur a Fadeev-Popov determinant
\begin{eqnarray}
J_{FP}={\det}_{\Omega^0}(-iL_{\sreeb}+iG_a)~,\nn
\end{eqnarray}
since the gauge transformation of $\Phi$ is $G_{\ep}\Phi=L_{\sreeb}\ep+i[\ep,\Phi]$. With $\Phi$ fixed, the rest of the fields contribute to a determinant factor as in (\ref{QZloc_toy_non_disct})
\begin{eqnarray}
\textrm{sdet}^{1/2}_{\Omega^1\oplus\Omega_H^{2+}}(-iL_{\sreeb}+iG_a)~,\nn
\end{eqnarray}
where $\Omega^1$ comes from $A$ and $\Omega_H^{2+}$ from $\chi$.
Combining this with the Fadeev-Popov determinant
\begin{eqnarray}
J_{vec}=\textrm{sdet}^{1/2}_{2\Omega^0\oplus\Omega^1\oplus\Omega_H^{2+}}(-iL_{\sreeb}+iG_a)~,\nn
\end{eqnarray}
but one needs to exclude from the zero forms their constant mode since these are not treated as gauge symmetry but as moduli of the Colomb branch. The hyper contribution is more straightforward, one uses the analogue of the toy model given in sec.\ref{QZsec_localisation}
\begin{eqnarray}
 J_{hyp}=\textrm{sdet}^{-1}_{\Omega_H^{0,\sbullet}}(-iL^s_{\sreeb}+iG_a)~,\nn
 \end{eqnarray}
where as a reminder $L_{\sreeb}^s$ is the spinor Lie derivative whose relation to the usual Lie derivative is given in (\ref{QZL_X_L_X^s}).

To evaluate the first determinant, the complex can be decomposed into
\begin{eqnarray}
 2\Omega^0\oplus\Omega^1\oplus\Omega_H^{2+}=\Omega^{0,0}\oplus\Omega^{0,1}_H\oplus \Omega^{0,2}_H\bigoplus c.c~,\nn
 \end{eqnarray}
so we just need to compute the determinant taken on the complex $\Omega^{0,\sbullet}_H$, which is the Kohn-Rossi complex with the differential $\bar\partial_H$ given in (\ref{QZbar_partial_H}). It will be explained in detail in sec.\ref{QZsec_Td} that the super-determinant cancels out totally except those that are in the $\bar\partial_H$ cohomology, leaving only
\begin{eqnarray} J_{vec}=\textrm{sdet}_{H^{0,\sbullet}_{\bar\partial_H}}(-iL_{\sreeb}+iG_a)~,\nn\\
J_{hyp}=\textrm{sdet}^{-1}_{H^{0,\sbullet}_{\bar\partial_H}}(-iL^s_{\sreeb}+iG_a)~.\nn\end{eqnarray}
For the vector multiplet, we needed to take a square root but we will ignore the possible phase, and we also remember that we exclude the constant mode.

Assemble the two determinants together with the classical action evaluated at (\ref{QZconfig_perturb}), and Wick rotate $a\to ia$
\begin{eqnarray}
Z^{pert}
=\int\limits_{\FR{su}}da~e^{-\frac{8\pi^3 r}{g_{YM}^2}\varrho\,\QZTr[a^2]}\cdotp
\frac{{\det}_{adj}' ~  \textrm{sdet}_{\Omega_H^{0,\sbullet}}(-iL_{\sreeb}-G_a)} {\det_{\underline{R}}\textrm{sdet}_{\Omega_H^{0,\sbullet}}(-iL^s_{\sreeb}-G_a)}\cdotp\frac{1}{{\det}'_{adj} (-G_a)}~,\nn\end{eqnarray}
in fact the last term  can be absorbed if one writes the integral of $a$ not over $\FR{su}(N)$ but over its cartan $\FR{t}$
\begin{eqnarray}
Z^{pert}
=\int\limits_{\FR{t}}da~e^{-\frac{8\pi^3 r}{g_{YM}^2}\varrho\,\QZTr[a^2]}\cdotp
\frac{{\det}_{adj}' ~  \textrm{sdet}_{\Omega_H^{0,\sbullet}}(-iL_{\sreeb}-G_a)} {\det_{\underline{R}}\textrm{sdet}_{\Omega_H^{0,\sbullet}}(-iL^s_{\sreeb}-G_a)}~.\label{QZZ_pert_text}\end{eqnarray}

\subsection{The determinant}\label{QZsec_Td}

This section makes heavy use of differential geometrical properties of the toric Sasaki-Einstein manifolds. We try to provide enough stepping stones in the text, further details can
 be found in the appendix.

From the travail of previous sections, the whole localisation reduces to the computation of the super determinant in (\ref{QZZ_pert_text})
\begin{eqnarray}
{s\det}_{\QZcal V}(-iL_{\sreeb}+x)~,~~~{\QZcal V}=\oplus\Omega_H^{0,\sbullet}\label{QZdet_kit}
\end{eqnarray}
taken over the horizontal anti-holomorphic forms ${\QZcal V}=\Omega_H^{0,\sbullet}$. Even though ${\QZcal V}$ is infinite dimensional, the presence of susy guarantees massive cancellation in the super determinant. The problem is to how to keep track of the cancellation, and more importantly, the remainders after the cancellation.

Let $P$ (respectively  $P_+$ and $P_-$) be the projectors that projects a 1-form to its horizontal (respectively
 horizontal hollomorphic and anti-holomorphic) components, defined in (\ref{QZprojectors}). We define an operator that acts on horizontal $(p,q)$ forms
\begin{eqnarray}
\bar\partial_H\omega=dx^r(P_-)_r^{~s}\nabla_s\omega+iq\kappa\wedge\omega,~~~\omega\in\Omega_H^{p,q}~.\label{QZKohn_Rossi}
\end{eqnarray}
One can check that it sends $\Omega_H^{p,q}\to\Omega_H^{p,q+1}$ and it is nilpotent, i.e. it is a differential of the complex $\Omega_H^{p,\sbullet}$. The cohomology of $\bar\partial_H$ is known as the \emph{Kohn-Rossi} cohomology (see \cite{QZKohnRossi}, we shall soon show that $\bar\partial_H$ is the restriction of the Dolbeault operator on the cone $C(M)$ to the boundary $M$, which is the setting of \cite{QZKohnRossi}). One can also couple $\bar\partial_H$ to the gauge connection, and all properties still hold if the gauge curvature is horizontal anti-self-dual; hence this operator is a differential at any instanton background.

The $\bar\partial_H$-complex is not elliptic: clearly the symbol of $\bar\partial_H$ is not exact along the $\reeb$ direction; thus its cohomology is of infinite dimension. But fortunately, for the toric Sasaki manifolds, we have a powerful index theorem that can handle the difficulty. For toric Sasaki geometry, the isometry contains $G=U(1)^3$, and the Reeb is a linear combination of the three $U(1)'s$. Then the $\bar\partial_H$-complex is an elliptic complex transverse to the $G$-action, since its symbol is elliptic transverse to the Reeb. Furthermore, $\bar\partial_H$ is  invariant under the $G$-action, since all structures appearing in (\ref{QZKohn_Rossi}), the metric, $\reeb$, $\kappa$ (and hence also $J$, since $J\sim d\kappa$) are invariant under the $G=U(1)^3$ isometry.
Then we have the decomposition of the $\bar\partial_H$ cohomology into the representations of $G$ (see Theorem  2.2 \cite{QZEllip_Ope_Cpct_Grp})
\begin{eqnarray}
H^{0,p}_{\bar\partial_H}=\bigoplus_i m^p_iR_i~,~~m^p_i\in\QZBB{Z}_{\geq0}~,\label{QZdecomp_char}
\end{eqnarray}
where the $m$'s are multiplicities of the representation of $R_i$. In the case $G=U(1)^3$, the representations are just labelled by 3 charges, which we will organize into an integer valued 3-vector.

The localisation technique for transversely elliptic operators due to Atiyah \cite{QZEllip_Ope_Cpct_Grp} allows us to compute the alternating differences of $m_i$. However this is a fairly involved task, so we present first a technique sketched by Schmude \cite{QZSchmude:2014lfa}.
\subsection{Schmude's approach}

The key observation is that the Dolbeault operator $\bar\partial^6$ on the 6d K\"ahler cone acts as
\begin{eqnarray} \bar\partial^6
=\frac12(t^{-1}dt-i\kappa)(L_{t\partial_t}+iL_{\sreeb})-\frac{i}{2}d\kappa\,\iota_{t\partial_t}+\bar\partial^5_H\label{QZbar_partial_6}~,\end{eqnarray}
where we have inserted $5,6$ to indicate whether an object is a 5d or 6d one.

Consider $H_{\bar\partial_H}^{0,\sbullet}$, and decompose it according to (\ref{QZdecomp_char}), meaning that we can discuss $H_{\bar\partial_H}^{0,\sbullet}$ assuming a fixed $U(1)^3$ charge.
 In all our considerations we assume the toric setup.
Take $\alpha$ a representative of $H^{0,\sbullet}_{\bar\partial_H}$, we can assume that it has charge vector $\vec q$, then its $L_{\sreeb}$ eigenvalue is $L_{\sreeb}\alpha=i(\vec\reeb\cdotp\vec q)\alpha$, where we have used an integer 3-vector $\vec\reeb$ to express the Reeb as a linear combination of the three $U(1)$ isometry. We can now extend $\alpha$ to the K\"ahler cone as
\begin{eqnarray}
\alpha\to\tilde\alpha=t^{\vec \sreeb\cdotp\vec q}\alpha~,\label{QZextension}
\end{eqnarray}
looking at (\ref{QZbar_partial_6}), $\tilde\alpha$ will be annihilated by $\bar\partial^6$. If $\vec\reeb$ is assumed to be within the dual cone (see sec.\ref{QZsec_TSm}) then $\vec\reeb\cdotp \vec q\geq0$ and so $\tilde\alpha$ is well-defined within the 6d K\"ahler cone.

On the other hand if $\alpha=\bar\partial_H \beta$, then
\begin{eqnarray}
 \tilde\alpha=\bar\partial^6\tilde\beta~,\nn
 \end{eqnarray}
where $\tilde\beta=t^{\vec \sreeb\cdotp\vec q}\beta$. Thus we have a well-defined map of cohomology
\begin{eqnarray}
H^{0,\sbullet}_{\bar\partial_H}(M)\to H_{\bar\partial^6}^{0,\sbullet}(C(M))~.\label{QZmap_cohom}
\end{eqnarray}
The left inverse to the extension map (\ref{QZextension}) is the restriction map that restricts a form to the surface $t=1$. This already shows that the restriction map is onto, while the extension map is into. From the injectivity we deduce
\begin{eqnarray} H^{0,1}_{\bar\partial_H}(M)=0\nn\end{eqnarray}
since $\pi_1(C(M))$ is at most torsion, and $C(M)$ is K\"ahler.

We turn now to $H^{0,0}_{\bar\partial_H}$. Since $H_{\bar\partial^6}^{0,0}(C(M))$ are the holomorphic functions on $C(M)$ and the latter has a very convenient description in the toric case: they correspond to integer lattice points in the cone $C$. One can read off the charges under $U(1)^3$ of the function from the coordinates of the lattice point that represents the function. This also shows that the restriction to $t=1$ is injective, since there is one unique holomorphic function with a given $U(1)^3$ charge, two functions with different charges cannot cancel each other when restricted to $t=1$. In this way we have a complete answer for (\ref{QZdecomp_char}) at degree zero and one.

For degree two, one can also show that (\ref{QZextension}) is an extension and compute $H^{0,2}_{\bar\partial^6}(C(M))$ using Serre duality. But we can in fact make the Serre-duality explicit here. Let $\bar\varrho\in \Omega_H^{0,2}$ satisfying $\partial_H\bar\varrho=0$ (from this we also have $d^{\dag}\bar\varrho=0$). Now any section of $\Omega^{0,2}_H$ is of the form $\bar f\bar\varrho$ for some function $\bar f$, we only need to sift out the $\bar\partial_H$-exact ones to get $H^{0,2}_{\bar\partial_H}$. On $(0,2)$ forms we have
\begin{eqnarray}
\bar\partial_H^{\dag}(\bar f\bar\varrho)=d^{\dag}(\bar f\bar\varrho)=-g^{pq}(\partial_p\bar f)\bar\varrho_{qr}dx^r~.\nn
\end{eqnarray}
The right hand side  is zero iff $\bar f$ satisfies $\partial_H\bar f=0$. Note that if $\bar\partial_H^{\dag}(\bar f\bar\varrho)=0$, then $\bar f\bar\varrho$ is orthogonal to any $\bar\partial_H$-exact forms, so we reach the conclusion that
\begin{eqnarray} H^{0,0}_{\bar\partial_H}(M)\stackrel{f\to \bar f\bar\varrho}{\longrightarrow} H^{0,2}_{\bar\partial_H}(M)\label{QZwedge_rho}\end{eqnarray}
is an isomorphism \footnote{In the proof, we have not treated some analytical issues carefully, such as how to define the Hilbert space where the horizontal forms reside, but this is slightly of the topic of the paper. The same omission was in the treatment of \cite{QZQiu:2014oqa}, but we believe that the our result will not be affected by these technicalities.}

Now we can wrap up the lengthy discussion and get the index. We introduce some formal variables $s_a,~a=1,2,3$ and use monomials of such to denote a representation of $U(1)^3$. For example
$s_1^2s_2^{-5}s_3$ is a representation of charge 2 under the 1st $U(1)$, charge $-5$ under the 2nd and charge 1 under the 3rd. Then the index
\begin{eqnarray} &&\textrm{ind}_{U(1)^3}\,\bar\partial_H=\sum_{\vec m\in C\cap \QZBB{Z}^3}{\vec s}^{\;\vec m}+\sum_{\;-\vec m\in C^{\circ}\cap\QZBB{Z}^3}\vec s^{\;-\vec m},\label{QZThe_index}\\
\textrm{where}&& {\vec s}^{\vec m}=s_1^{m_1}s_2^{m_2}s_3^{m_3},\nn\\
&&C=\{\vec r\in\QZBB{R}^3,~~\vec r\cdotp \vec v_i\geq0\},\nn\\
&&C^{\circ}=\{\vec r\in\QZBB{R}^3,~~\vec r\cdotp \vec v_i>0\},\nn\end{eqnarray}
where $i$ runs over all faces of the cone.
The first summand comes from $H^{0,0}_{\bar\partial_H}$ and is straightforward. For the second term, since we see from (\ref{QZwedge_rho}) that $H^{0,2}_{\bar\partial_H}$ are also 1-1 to lattice points in the cone, but the $U(1)^3$-charge is reversed and then shifted by the charge of $\bar\varrho$. So we should have written the second summand as $\sum_{\;-\vec m\in C\cap\QZBB{Z}^3}\vec s^{\;-\vec m-\vec\xi}$
where $-\vec\xi$ is the charge vector for $\bar\varrho$. But if we use the 1-Gorenstein condition condition $\vec\xi\cdotp\vec v_i=1,~\forall i$ (see the discussion around (\ref{QZ1-gorenstein}), we can write the sum as in (\ref{QZThe_index}). But note that (\ref{QZThe_index}) is valid even for toric Sasaki manifolds, i.e. when $\bar\varrho$ does not exist.

We continue our computation of the determinant. From the index, we read off
\begin{eqnarray} \textrm{sdet}_{\Omega_H^{0,\sbullet}}(-iL_{\sreeb}+x)=\prod_{\vec n\in C\cap \QZBB{Z}^3}\big(\vec n\cdotp \vec \reeb+x\big)\big(-\vec n\cdotp \vec \reeb-2f_{\sreeb}+x\big),\label{QZdet_result}\end{eqnarray}
where the first product comes from $H^{0,0}_{\bar\partial_H}$ and $\vec n\cdotp \vec\reeb$ is the $-iL_{\sreeb}$-eigenvalue; the second term comes from $H^{0,2}_{\bar\partial_H}$ and their $-iL_{\sreeb}$ eigenvalue has been explained in the last paragraph.
%To see this, a lattice point $\vec n$ correspond to charge $[n^1,n^2,n^3]$ under the three $U(1)$'s, while the Reeb is a linear combination of these same $U(1)$'s, with components $\reeb^{1,2,3}$. The eigenvalue is thus $\vec n\cdotp \vec\reeb$ for $H_{\bar\partial_H}^{0,0}$. For $H_{\bar\partial_H}^{0,2}$, its charge can be deduced from the pairing (\ref{QZpair_KR}) to be $-\vec n\cdotp \vec\reeb-2f_{\sreeb}$.
%{\color{blue}end incomplete}

\begin{QZexample}
As a more familiar example, take $M=S^5$. Then the cone is just the first octant $C=\QZBB{R}^3_{\geq0}$. Take also $\vec\reeb=[1,1,1]$, then $f_{\sreeb}=3$ and
\begin{eqnarray} \textrm{sdet}=\prod_{m\geq0}\big(m+x\big)^{(m+1)(m+2)/2}\big(-m-3+x\big)^{(m+1)(m+2)/2}~.\label{QZdet_S^5}\end{eqnarray}
The multiplicity $(m+1)(m+2)/2$ comes as follows. Fixing the plane $\vec n\cdotp\vec\reeb=m$, then its intersection with the cone contains $(m+1)(m+2)/2$ lattice points.

In fact, with the Reeb given by its charge $[1,1,1]$, it corresponds to the $U(1)$ vector field in the Hopf fibration $U(1)\to S^5\to \QZBB{C}P^2$. In this case one can compute the $\bar\partial_H$-cohomology using a 'Fourier transform'. For example computing $H^{0,0}_{\bar\partial_H}(S^5)$ with fixed $(-iL_{\sreeb})$-eigenvalue $m$ amounts to computing
$H^0(\QZBB{C}P^2,{\QZcal O}(m))$, and the answer is $(m+1)(m+2)/2$ for $m\geq0$ (which is the number of monomials homogeneous of degree $m$ in three variables) and zero otherwise. In general
\begin{eqnarray} \dim H^{0,0}(\QZBB{C}P^2,{\QZcal O}(n))&=&\bigg\{\begin{array}{cc}
                                           \frac{1}{2}(n+1)(n+2) & n\geq0 \\
                                           0 & n<0
                                         \end{array},\nn\\
     \dim H^{0,1}(\QZBB{C}P^2,{\QZcal O}(n))&=&0\nn\\
     \dim H^{0,2}(\QZBB{C}P^2,{\QZcal O}(n))&=&\bigg\{\begin{array}{cc}
                                           \frac{1}{2}(n+1)(n+2) & n\leq-3 \\
                                           0 & n>-3
                                         \end{array}\label{QZcoho_P2}.\end{eqnarray}
%The degree 1 cohomology $H^1(\QZBB{P}^2,{\QZcal O}(m))=0$ for all $m$.
The group $H^{0,2}_{\bar\partial_H}(S^5)\sim H^2(\QZBB{C}P^2,{\QZcal O}(m))\sim H^0(\QZBB{C}P^2,{\QZcal O}(-m-3))^*$, where the last duality is the Serre duality and takes the place of (\ref{QZwedge_rho}).
After a change of summation variable, we gets the second exponential in (\ref{QZdet_S^5}).
\end{QZexample}

\subsection{Generalised multiple sine}

Up to an overall sign, the product of (\ref{QZdet_result}) is an interesting generalisation of the multiple sine functions. Recall that the usual multiple sine function is defined as
\begin{eqnarray}
S_r(x|\omega)=\prod_{m_a\geq0}\big(\sum_{a=1}^rm_a\omega_a+x)\prod_{m_a>0}\big(\sum_{a=1}^rm_a\omega_a-x)^{(-1)^{r-1}}~.\label{QZmultiple_sine}
\end{eqnarray}
One can define a generalised version of multiple sines associated with a cone in $\QZBB{R}^r$,
\begin{eqnarray}
S_r^C=\prod_{\vec m\in C\cap\QZBB{Z}^r}(\vec \omega\cdotp\vec m+x)\prod_{\vec m\in C^{\circ}\cap\QZBB{Z}^r}(\vec \omega\cdotp\vec m-x)^{(-1)^{r-1}}~.\label{QZgen_multiple_sine}
\end{eqnarray}
So the standard multiple sine corresponds to the cone that is the first orthant of $\QZBB{R}^r$.
For more properties of (generalised) multiple sines, see
\cite{QZMR2101221} and \cite{QZTW}

For our problem, we have the cone $C\subset\QZBB{R}^3$ which is also the image of the moment map of $C(M)$. So (\ref{QZdet_result}) can be written as the generalised triple sine function associated with this cone
\begin{eqnarray} (\ref{QZdet_result})\sim
S_3^C (x| \vec\reeb)=\prod\limits_{\vec m\in C(X)\cap\QZBB{Z}^3}\big(\vec m\cdotp\vec \reeb+x\big)\prod\limits_{\vec m\in C^{\circ}(X)\cap\QZBB{Z}^3}\big(\vec m\cdotp\vec \reeb-x\big)~.\nn
\end{eqnarray}

We mainly focus on the case when $M$ is simply connected SE, then $C$ is 1-Gorenstein. As we saw, one has
$\vec\xi\in\QZBB{Z}^3$ such that $\vec\xi\cdotp\vec v_i=1$ and the second product can be written as
$\prod\limits_{\vec m\in C(X)\cap\QZBB{Z}^3}\big(\vec m\cdotp\vec \reeb+\vec\xi\cdotp\vec\reeb-x\big)$ and $\vec\xi\cdotp\vec\reeb$ is precisely the shift $2f_{\sreeb}$.

So we have finished our localisation computation for the zero instanton sector
\begin{eqnarray}
Z^{pert}
=\int\limits_{\FR{t}}da~e^{-\frac{8\pi^3 r}{g_{YM}^2}\varrho\,\QZTr[a^2]}\cdotp
\frac{{\det}_{adj}' ~  S_3^C(ia| \vec\reeb)}{\det_{\underline{R}}S_3^C(ia+im+\vec\xi\cdotp\vec\reeb/2| \vec\reeb)}~,\label{QZZ_pert_fin}\end{eqnarray}
where a mass $m$ is generated for the hyper-multiplet by a simple shift of $a$, i.e. $m$ is regarded as a background gauge connection coupled to the hyper. The matrix model (\ref{QZZ_pert_fin}) is discussed further in \volcite{MI}. 

\subsection{Conjecture for the full answer}

The answer (\ref{QZZ_pert_fin}) corresponds to the contribution of the trivial connection. In order to derive the full answer we have to analyze the contact instantons (\ref{QZcontact-instanton}) and
 perform the one-loop calculations over every non-trivial solutions. As it stands the problem is hard to solve from the first principles. However it is natural to expect that
  only the configurations invariant under full $U(1)^3$ action will contribute to the integral. The invariant configurations will tend to localise around the close Reeb orbits (for the generic
   choice of ${\reeb}$ there will be only a few closed orbits and they are called Reeb orbits). Thus around every Reeb orbit the complex and the calculation will boil down to
    the calculation on $\QZBB{C}^2 \times S^1$, very much in analogy with Pestun's calculation  \cite{QZPestun:2007rz} on $S^4$ and its reduction to $\mathbb{C}^2$.  In order to
     conjecture the full answer we need to identify the parameters on toric SE manifold with with the parameters of Nekrasov's instanton partition function on $\QZBB{C}^2 \times S^1$ corresponding
      to each closed Reeb orbit. This can be done either geometrically or by studying the factorisation properties of the perturbative answer \cite{QZQiu:2014oqa}.
     The full answer is written as
 \begin{eqnarray}
Z^{full}
=\int\limits_{\FR{t}}da~e^{-\frac{8\pi^3 r}{g_{YM}^2}\varrho\,\QZTr[a^2]}\cdotp
\frac{{\det}_{adj}' ~  S_3^C(ia| \vec\reeb)}{\det_{\underline{R}}S_3^C(ia+im+\vec\xi\cdotp\vec\reeb/2| \vec\reeb)} \prod\limits_{i=1}^n Z_{\QZBB{C}^2 \times S^1}^{\rm Nekrasov} (\beta_i, \epsilon_i, \epsilon'_i)~,\label{QZZ_pertfull}\end{eqnarray}
 where $\beta_i$ is radius of $S^1$, $\epsilon_i, \epsilon'_i$ are equivariant parameters on $\QZBB{C}^2$. Here $n$ is the number of closed Reeb orbits and the parameters
 $\beta_i , \epsilon_i, \epsilon'_i$ can be read off from the toric data \cite{QZQiu:2014oqa}. This conjecture is discussed further in \volcite{PA}.

\section{Asymptotics and comparison with flat space}\label{QZsec_Aacwfs}

In this section, we will compare our result with a one-loop flat space computation, in particular, we will obtain a match between the precise coefficient of the effective ${\QZcal A}^3$ term generated at 1-loop.

We first analyze the large $x$ behaviour of the generalised triple sine functions. The process is a bit technical and we start from a toy model
\begin{eqnarray} S_1=\prod_{n=0}^{\infty}(x+n\omega)\cdotp\prod_{n=0}^{\infty}(\omega-x+n\omega)=2\sin\frac{\pi x}{\omega}~.\nn\end{eqnarray}
 Here we assume that all infinite product are regularized.
Take the first product and write it using the zeta function regularisation
\begin{eqnarray} \log\prod_{n=0}^{\infty}(x+n\omega)&=&-\frac{\partial}{\partial s}\frac{1}{\Gamma(s)}\int_0^{\infty}\sum_{n=0}^{\infty}e^{-(n\omega+x)t}t^{s-1}dt\Big|_{s=0}\nn\\
&=&-\frac{\partial}{\partial s}\frac{1}{\Gamma(s)}\int_0^{\infty}\frac{e^{-xt}}{1-e^{-\omega t}}t^{s-1}dt\Big|_{s=0}~,\nn\end{eqnarray}
where it is assumed $\re\omega>\re x>0$ and we are interested in $\im x\gg 0$. In this regime, one can replace $(1-e^{-\omega t})^{-1}$ with its Laurent expansion at $t=0$, the error is of order $x^{-1}$ (see sec.6.1 of \cite{QZQiu:2013pta}). Furthermore since $\lim_{s\to0}\Gamma(s)^{-1}\to 0$, one needs only keep the singular terms from the integral. Thus one keeps only terms of order $t^{-1},~t^0$
in the previous Laurent series. These terms will produce $\Gamma(s-1),~\Gamma(s)$ and therefore survive the limit $s\to 0$. The net result is
\begin{eqnarray} \log\prod_{n=0}^{\infty}(x+n\omega)=-\frac{1}{\omega}(x\log x-x)-\frac12\log x+{\QZcal O}(x^{-1})~.\nn\end{eqnarray}
One replaces $x$ with $\omega-x$ in the second product, and in total one gets
\begin{eqnarray} \log S_1(x|\omega)=\frac{\pi}{\omega}|\im x|-\frac{i\pi}{2}\mathrm{sgn}\,(\im x)+{\QZcal O}(x^{-1})~.\nn\end{eqnarray}

In the higher dimension case, we have the following formula that expresses the asymptotic behaviour of a generalised triple sine in terms of the geometrical data of the cone (we assume that the cone is 1-Gorenstein)
\begin{eqnarray} \log S_3^C(x|\vec\reeb)\sim -i\pi\textrm{sgn}(\im x)\Big(\big(\frac{x^3}{3\reeb^1}+\frac{\reeb^1x}{6}\big)\sum_i\frac{4}{|v_i|}A_i
+\frac{x}{12}\frac{1}{2\pi}\sum_i\beta_i\Big)=-V_{vec}(x)~,\label{QZasymp_v}\\
\log S_3^C(x+\frac12\vec\xi\cdotp\vec\reeb|\vec\reeb)\sim -i\pi\textrm{sgn}(\im x)\Big(\big(\frac{x^3}{3\reeb^1}-\frac{\reeb^1x}{12}\big)\sum_i\frac{4}{|v_i|}A_i
+\frac{x}{12}\frac{1}{2\pi}\sum_i\beta_i\Big)=V_{hyp}(x)~.\label{QZasymp_h}\end{eqnarray}
In the above one should understand $x$ as $x^it_i$ for some basis $\{t_i\}$ of the Lie algebra, and $\im x$ takes the imaginary part of each $x^i$. The rest of the term in this asymptotic formula can all be read off from the geometry of the cone. The $\beta_i$ are the length of the closed Reeb orbits
\begin{eqnarray} \beta_i=\frac{2\pi}{\det[\vec v_i,\vec v_{i+1},\vec\reeb]},\nn\end{eqnarray}
where $\vec v_i,\vec v_{i+1}$ are the normals of the two faces that intersect at the $i^{th}$ edge of the cone.
For the $A_i$'s, let us cut the cone off with the plane $\vec y\cdotp \reeb=1/2,~\vec y\in\QZBB{R}^3$, then $A_i$ is the area of the $i^{th}$ face.

To summarise, asymptotically, the matrix model integral is given by
\begin{eqnarray}
Z^{pert}\sim \int\limits_{\FR{t}}da~e^{-\frac{8\pi^3 r}{\gYM^2}\varrho\,\QZTr[a^2]}\cdotp
\exp\big(-\QZTr_{ad}V_{vec}(iar)-\QZTr_{\underline{R}}V_{hyp}(iar)\big)~,\label{QZZ_asymp}\end{eqnarray}
with $V_{vec,hyp}$ given in (\ref{QZasymp_v}) and  (\ref{QZasymp_h}), and $r$ is of dimension length that controls the size of the manifold.

\subsection{Comparison with flat space}
In particular, we can consider the $S^5$ case, where the cone is the standard one. Then $A_i=1/8$, $\reeb^1=\sum \omega_i$, and $\beta_i=2\pi \omega_i^{-1},~i=1,2,3$. If the sphere is the round one all $\omega_i=1$, we get then the effective potentials
\begin{eqnarray} V^{S^5}_{vec}(x)\sim i\pi\textrm{sgn}(\im x)\big(\frac{x^3}{6}+x\big)~,\nn\\
V^{S^5}_{hyp}(x)\sim -i\pi\textrm{sgn}(\im x)\big(\frac{x^3}{6}-\frac{x}{8}\big)~.\label{QZsymp_S^5}\end{eqnarray}

\begin{figure}[h]
\begin{center}
\begin{tikzpicture}[scale=0.7,node distance=1cm, auto]
 %nodes
 \node[punkt] (13) {placed on $S^5$};
 \node[punkt, left=of 13] (12) {One loop generates ${\QZcal A}^3$ term}
   edge[pil] (13.west);
 \node[punkt, left=of 12] (11) {SYM on flat space w/o ${\QZcal A}^3$ term}
   edge[pil] (12.west);
 \node[punkt, below= of 11] (21) {SYM on $S^5$}
   edge[pil,<-] (11.south);

 \node[punkt, right=of 13] (14) {evaluated in Coulomb branch}
   edge[pil,<-] (13.east);
 \node[punkt, below= of 12] (22) {Localisation}
   edge[pil,<-] (21.east);
%   edge[pil,->,bend right=10] (14.south);
 \node[punkt, below= of 14] (24) {$\sigma^3/r^2$ term}
   edge[pil,<-] (14.south);
 \node[punkt, right= of 22] (23) {large radius limit}
   edge[pil,<-] (22.east)
   edge[pil,->] (24.west);
\end{tikzpicture}
\end{center}\caption{One starts from super Yang-Mills on flat space without the ${\QZcal A}^3$ term in the prepotential. The one loop contribution generates an ${\QZcal A}^3$ term, then the whole system can still be put on $S^5$ and evaluated in Coulomb branch and will produce $\sigma^3$ term. On the other hand, one can first place the theory on $S^5$, perform localisation, take $r\to\infty$ limit and obtain the $\sigma^3/r^2$ term.}\label{QZfig_chart}
\end{figure}
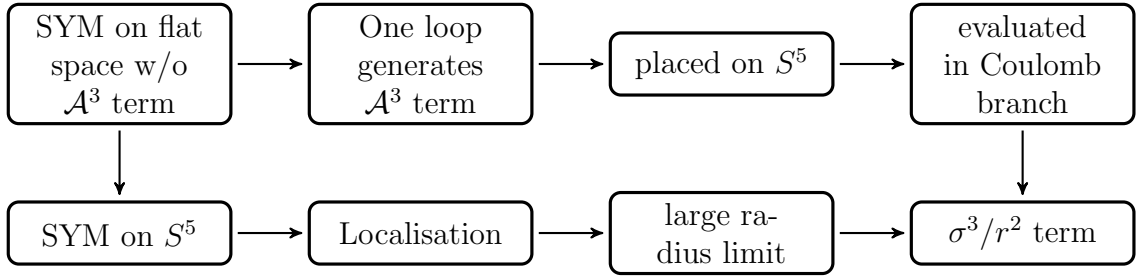

The relevance of the asymptotic bevahiour is that it controls the flat space limit. If one restores the dimensionful parameter $r$, which is the radius of $S^5$, we obtain the effective action
\begin{eqnarray} S_{eff}=\frac{8\pi^3 r^3}{\gYM^2}\,\QZTr_f[a^2]+\sum_{\alpha\in\textrm{roots}}\big(\frac{r^3}{6}|\QZbra a,\alpha\QZket|^3-r|\QZbra a,\alpha\QZket|\big)-\sum_{\mu\in\textrm{weights}}\big(\frac{r^3}{6}\QZbra a,\mu\QZket|^3+\frac{r}{8}\QZbra a,\mu\QZket\big)~.\label{QZZ_asymp}\end{eqnarray}
From this one notices that since the volume of $S^5$ is $\pi^3 r^5$, the effective potential is suppressed by $r^{-2}$
\begin{eqnarray} V_{eff}=\frac{8}{r^2\gYM^2}\,\QZTr_f[a^2]+\sum_{\alpha\in\textrm{roots}}\frac{1}{6r^2}|\QZbra a,\alpha\QZket|^3-\sum_{\mu\in\textrm{weights}}\frac{1}{6r^2}|\QZbra a,\mu\QZket|^3+{\QZcal O}(r^{-4}),\label{QZV_asymp}\end{eqnarray}
i.e. $V_{eff}$ computed in the Colomb branch vanishes as $r\to\infty$ and the $V_{eff}$ we have above is due entirely to the curved space effect. But this is not surprising, since $a$ is the bottom component of the superfield ${\QZcal A}$, and if the only nonzero background of ${\QZcal A}$ is $a=const$, then nothing will survive the superspace integral. Therefore the comparison with the flat space computation will take an indirect route. The comparison goes as the following chart \ref{QZfig_chart}.

\subsection{The 1-loop effective action}
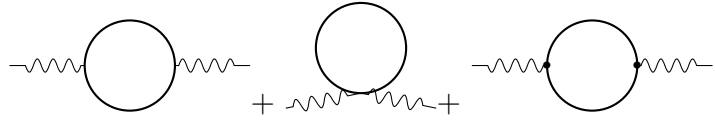
\begin{figure}[h]
\begin{center}
\begin{tikzpicture}[scale=1]
 \draw[thick] (0cm,0cm) circle(.6cm);
 \draw[snake=coil, line before snake=.5mm, line after snake=1mm, segment aspect=0, segment length=6pt, color=black] (.6,0cm) -- (1.6,0cm);
 \draw[snake=coil, line before snake=.5mm, line after snake=1mm, segment aspect=0, segment length=6pt, color=black] (-.6,0cm) -- (-1.6,0cm);
\end{tikzpicture}+
\begin{tikzpicture}[scale=1]
 \draw[thick] (0cm,0cm) circle(.6cm);
 \draw[snake=coil, line before snake=1mm, line after snake=1mm, segment aspect=0, segment length=6pt, color=black] (-1,-.8cm) -- (0,-.6cm) -- (1,-.8cm);
\end{tikzpicture}+
\begin{tikzpicture}[scale=1]
 \draw[thick] (0cm,0cm) circle(.6cm);
 \draw[snake=coil, line before snake=.5mm, line after snake=1mm, segment aspect=0, segment length=6pt, color=black] (.6,0cm) -- (1.6,0cm);
 \draw[snake=coil, line before snake=.5mm, line after snake=1mm, segment aspect=0, segment length=6pt, color=black] (-.6,0cm) -- (-1.6,0cm);
 \node at (-.6,0) {$\sbullet$};\node at (.6,0) {$\sbullet$};
\end{tikzpicture}\caption{The 1-loop diagrams. In the third diagram $\bullet$ represents the insertion $F_{pq}J^{pq}$. And the momentum in the loop is to run clockwise.}
\label{QZfig_1-loop}
\end{center}
\end{figure}

Consider the flat space action (\ref{QZaction_flat}) and we will compute the 1-loop effective action at some background. The computation is standard (see for example \cite{QZPeskinSchroeder}),
one just needs to compute the determinant
\begin{eqnarray} S_{eff}(\phi_{cl})-S_{0}(\phi_{cl})=\textrm{sTr}\,\frac12\log \frac{\partial^2 S_0(\phi)}{\partial \phi\partial \phi}\Big|_{\phi=\phi_{cl}}.\label{QZ1-loop_det}\end{eqnarray}
It is also easier to consider its 6d lift of the 5d action, for which $\sigma$ becomes the temporal component of the gauge field, and hence we just have the gauge fields, fermions plus the ghosts.
Split the gauge field as $A={\sf A}+a$, with ${\sf A}$ some background, denote by ${\sf D}$ the covariant derivative with ${\sf A}$ and ${\sf F}$ its curvature.
For fields of different spins we have a uniform description of the quadratic term
\begin{eqnarray} S''(\phi_{cl})=-{\sf D}^2+{\sf F}_{mn}J^{mn},\nn\end{eqnarray}
where $J$ is the angular momentum generator
\begin{eqnarray} &&J^{mn}=0~~\textrm{spin }0;~~~~J^{mn}=\frac{i}{2}\Gamma^{mn} ~~\textrm{spin }1/2,~~~~(J^{mn})_{pq}=i\delta^{mn}_{pq}~~\textrm{spin }1,\nn\\
&&\hspace{3cm}\QZTr[J^{rs}J^{pq}]=C(j)g^{r[p}g^{q]s}\label{QZCasimir_spin}\end{eqnarray}
with $C(1)=2$ and $C(1/2)=1$ for the last two cases.
The determinant (\ref{QZ1-loop_det}) reduces to
\begin{eqnarray} \det \Delta_{r,j}=\det(-\partial^2+\Delta^1+\Delta^2+\Delta^J),\nn\end{eqnarray}
for each field of representation $r$ and spin $j$. The various terms read
\begin{eqnarray} \Delta^1=i(\partial^m{\sf A}_m+{\sf A}_m\partial^m),~~~\Delta^2={\sf A}^m{\sf A}_m,~~~\Delta^J={\sf F}_{pq}J^{pq}.\nn\end{eqnarray}
Out of the computation we aim to get the coefficient of the term $\sigma F*F$, so we choose a convenient (supersymmetric) background ${\sf A}^{1-5}\in\FR{h}=\textrm{Lie}\,H$, and $\sigma={\sf A}^0\in\FR{h}$ a constant. We compute the determinant up to second order in ${\sf A}^{1-5}$, so the relevant diagrams are in fig.\ref{QZfig_1-loop}.
The first two diagrams give
\begin{eqnarray}&& I+II=-\frac12\,d(j)\sum_{\mu\in \textrm{wght}}\int\frac{d^dk}{(2\pi)^d}\QZbra\mu|{\sf A}_m(k){\sf A}_n(-k)|\mu\QZket\frac{\Gamma(2-d/2)}{(4\pi)^{d/2}}(k^2g^{mn}-k^mk^n)\nn\\
&&\hspace{4.2cm}\int_0^1 dx~(1-4x^2)(x(1-x)k^2+\QZbra\mu|\sigma|\mu\QZket^2)^{d/2-2},\label{QZdiagram_1_2}\end{eqnarray}
where $d(j)$ is the dimension of the spin $j$ representation.
The third diagram gives
\begin{eqnarray} &&III=-2C(j)\sum_{\mu\in \textrm{wght}}\int\frac{d^dk}{(2\pi)^d}\QZbra\mu|{\sf A}_m(k){\sf A}_n(-k)|\mu\QZket(k^2g_{mn}-k_mk_n)\frac{\Gamma(2-d/2)}{(4\pi)^{d/2}}\nn\\
&&\hspace{3.6cm}\int_0^1dx~(x(1-x)k^2+\QZbra\mu|\sigma|\mu\QZket^2)^{d/2-2}\label{QZdiagram_3},\end{eqnarray}
with $C(j)$ being the Casimir defined in (\ref{QZCasimir_spin}).

From the three contributions, one can extract the term $\sigma {\sf F}*{\sf F}$,
for a spin $j$ field of representation $r$, one gets
\begin{eqnarray} X_{r,j}=(I+II)+III&=&\sum_{\mu\in\textrm{wght}}~\QZbra\mu|{\sf F}*{\sf F}|\mu\QZket\frac{\Gamma(2-d/2)}{(4\pi)^{d/2}}\Big(\frac16 d(j)-2C(j)\Big)|\QZbra\mu|\sigma|\mu\QZket^2|^{d/2-2}\nn\\
&=&-\frac{1}{16\pi^2}\Big(\frac{1}{6}d(j)-2C(j)\Big)\QZTr\big[{\sf F}*{\sf F}|\sigma|\big].\nn\end{eqnarray}
Adding up the field content for the vector multiplet ($n_f=1$ is the number of Dirac fermions in 5d, the $1/2$ is because $\det\Delta_{ad,1/2}$ computes the determinant of $\slashed{D}^2$)
\begin{eqnarray} X_{ad,1}^{-1/2}+X_{ad,0}+X_{ad,1/2}^{n_f/2}+X_{ad,0}^{-1/2}=-\frac{1}{16\pi^2}\QZTr_{ad}\big[{\sf F}*{\sf F}|\sigma|\big],\nn\end{eqnarray}
and for the hyper-multiplet
\begin{eqnarray} X_{r_h,1/2}^{1/2}+X_{r_h,0}^{-4/2}=\frac{1}{16\pi^2}\QZTr_{r_h}\big[{\sf F}*{\sf F}|\sigma-m|\big].\nn\end{eqnarray}
Thus the effective potential is
\begin{eqnarray} \frac{1}{V}S_{eff}({\sf A})=\frac{1}{2g^2}\sum_a{\sf F}^a*{\sf F}^a+\frac{1}{16\pi^2}\QZTr_{ad}\big[{\sf F}*{\sf F}|\sigma|\big]-\frac{1}{16\pi^2}\QZTr_{r_h}\big[{\sf F}*{\sf F}|\sigma-m|\big].\label{QZVeff_1_loop}\end{eqnarray}
The $\sigma F*F$ term appearing above comes from the ${\QZcal A}^3$ term in the prepotential. Recall (\ref{QZA^3_term}) that from $c/6{\QZcal A}^3$, one gets in the action
\begin{eqnarray} \frac{c}{6}{\QZcal A}^3\to \frac{c}{2\pi^2}\QZTr\Big[\sigma\int F\wedge*F+(D\sigma)*(D\sigma)+\cdots\Big]+CS(5).\nn\end{eqnarray}
Also in passing from $\QZBB{R}^5$ to $S^5$, (\ref{QZA^3_term}) undergoes the change
\begin{eqnarray}\int \sqrt{g}d^6x\QZTr[\sigma(D\sigma)^2]\to \int \sqrt{g}d^5x\,\QZTr[\sigma\big(D\sigma^2+\frac{R}{12}\sigma^2+\frac{1}{r^2}\sigma^2\big)].\nn\end{eqnarray}
The role of $R\sigma^2/12$ term is clear, it is there to make $D\sigma^2$ term conformal invariant \footnote{Under an infinitesimal conformal transform $\delta g_{ij}=2g_{ij}\phi$, one has $\delta\sigma =\sigma\phi(2-d)/3$ and $\delta R=2(1-d)\square\phi-2\phi R$.}. As for the last $\sigma^3/r^2$ term, it comes from the $D^2$ term in (\ref{QZaction_vector}), by plugging in the expectation value $D_{IJ}=-2r^{-1}\sigma t_{IJ}$ in (\ref{QZaction_flat}), one gets an extra $\sigma^3/r^2$. Now use $R=20/r^2$, one gets the association
\begin{eqnarray} \sigma F\wedge *F~\sim~ \frac{8}{3r^2}\sigma^3\nn\end{eqnarray}
Thus from the effective potential (\ref{QZVeff_1_loop}), one gets on $S^5$ a potential term
\begin{eqnarray} \frac{1}{16\pi^2}\frac{8}{3r^2}\big(\sum_{\beta\in\textrm{root}}~|\QZbra\sigma,\beta\QZket|^3
-\sum_{\mu\in\textrm{wght}}~|\QZbra\sigma,\mu\QZket-m|^3\big).\nn\end{eqnarray}
This matches perfectly with (\ref{QZV_asymp}).

%
%From this we can read off the quantum corrected prepotential
%\begin{eqnarray} {\QZcal F}=\frac{1}{4g^2}\QZTr_f[\sigma^2]+\frac{1}{192\pi^2}\Big(\sum_{\beta\in\textrm{root}}~|\QZbra\sigma,\beta\QZket|^3-\sum_{\beta\in\textrm{wght}}~|\QZbra\sigma,\beta\QZket-m|^3\Big)\nn\\
%16\pi{\QZcal F}=\frac{4\pi}{g^2}\QZTr_f[\sigma^2]+\frac{1}{12\pi}\Big(\sum_{\beta\in\textrm{root}}~|\QZbra\sigma,\beta\QZket|^3-\sum_{\beta\in\textrm{wght}}~|\QZbra\sigma,\beta\QZket-m|^3\Big)\nn.\end{eqnarray}
%

%\begin{eqnarray} \Big(\frac{d(j)}{16}-C(j)\Big)\frac{-i}{2(32\pi)}\QZTr\int\frac{d^5k}{(2\pi)^5}(k^2g^{mn}-k^mk^n)(k^2)^{1/2}A_mA_n.\nn\end{eqnarray}
%\begin{eqnarray} -S_{eff}({\sf A})&=&\frac{1}{2g^2}\QZTr_f[F^2]-\frac{1}{2g^2}\QZTr_f[F^2]\frac{|k|}{64\pi C_2(r_f)}\Big(-\frac12(\frac{5}{16}-2)C_2(ad)+(\frac{1}{16})C_2(ad)\nn\\
%&&+\frac{n_f}{2}(\frac{4}{16}-1)C_2(r_f)-\frac{n_s}{2}(\frac1{16})C_2(r_s)\Big)\nn\end{eqnarray}
%In particular, for the $N=1$ theory with $n_{hyp}$ hyper
%\begin{eqnarray} \frac{1}{g(k)^2}&=&\frac{1}{g^2}-\frac{|k|}{64\pi C_2(r_f)}C_2(ad)\Big(-\frac12(\frac{5}{16}-2)+(\frac{1}{16})+\frac{1}{2}(\frac{4}{16}-1)-\frac{1}{2}(\frac1{16})
%+n_{hyp}\big(\frac{1}{2}(\frac{4}{16}-1)-\frac{4}{2}\frac{1}{16}\big)C_2(r_h)\Big)\nn\\
%&=&\frac{1}{g^2}-\frac{|k|}{32\pi}\big(\frac12C_2(ad)-\frac{n_{hyp}}{2}C_2(r_h)\big).\nn\end{eqnarray}

\bigskip
{\bf Acknowledgements} We thank  Tobias Ekholm, Johan K\"all\'en, Joseph Minahan, Anton Nedelin, Vasily Pestun,  Luigi Tizzano and
 Jacob Winding  for discussions and for the collaborations on this and related subjects.
The research of J.Q. is supported in part by the Max-Planck Institute for Mathematics in Bonn. The research of M.Z. is supported in part by
Vetenskapsr{\aa}det under grant \#2014-5517, by the STINT grant and by the grant ''Geometry and Physics"
 from the Knut and Alice Wallenberg foundation.

%\appendix

\section{Appendix. Geometrical setting}\label{QZsec_GS}
\subsection{Some basics of contact geometry}
\noindent $\bullet$ Contact structure. A contact structure on a 5-manifold $M$ is a smooth distribution of contact element $\xi$, which is a 4d subspace of the tangent space $TM$. This distribution is required to be non-integrable, and in this review $\xi$ will be called the transverse or horizontal plane.
If $\xi$ is given by the kernel of a 1-form $\kappa$, then the non-integrability says that $\kappa(d\kappa)^2\neq0$ everywhere. Note that this condition implies $d\kappa$ is non-degenerate on $\xi$, and serves as an analogue of the symplectic structure. Also from the same condition one has a unique vector field $\reeb$ called the Reeb vector field such that
\begin{eqnarray}
\iota_{\sreeb}\kappa=1~,~~\iota_{\sreeb}d\kappa=0~.\label{QZdef_Reeb}
\end{eqnarray}
One can split $TM$ into vertical and horizontal components using the projector
\begin{eqnarray}
 P=1-\reeb\otimes\kappa~.\nn
 \end{eqnarray}

\noindent $\bullet$ Contact metric structure.
In analogy with the symplectic case, one can construct purely algebraically an endomorphism $J:~\xi\to \xi$ and $J^2=-1$. The triple $(\xi,\kappa,J)$ is said to be a contact metric structure on $M$ if $J$ is compatible with $d\kappa$ in the sense that $1/2d\kappa J$ is a metric for $\xi$.
We also extend $J$ to an endomorphism of the entire $TM$ by defining its action on $\reeb$ as zero $J\reeb=0$, leading to
\begin{eqnarray}
J^2=-P=-1+{\reeb}\otimes\kappa~.\nn
\end{eqnarray}
One can write down a metric of the tangent bundle as the direct sum of the one on $\xi=\ker\kappa$ and the one along ${\reeb}$
\begin{eqnarray}
 g=\frac1{2}d\kappa J+\kappa\otimes\kappa~.\label{QZstd_metric}
 \end{eqnarray}
As a consequence
\begin{eqnarray}
&&g(JX,JY)=g(X,Y)-\kappa(X)\kappa(Y)~,\nn\\
&&d\kappa=-2gJ~,\label{QZkappa_J}\\
&&{\reeb}=g^{-1}\kappa~.\nn\end{eqnarray}
\begin{QZremark}
As a note of the general convention of the review, we do not make any distinction of $J$ when it serves as an endomorphism of $TM$, or of $T^*M$ or a 2-form on $M$, all of which are related by raising or lower an appropriate index with the metric.
\end{QZremark}
Let us fix the volume form of $M$ as\footnote{We remind the reader that the choice of volume form in \cite{QZKallen:2012va} is minus the current one.}
\begin{eqnarray}
 \textrm{vol}=\frac12\kappa\wedge J\wedge J=\frac18\kappa\wedge d\kappa\wedge d\kappa~,\label{QZvolume_form}
 \end{eqnarray}
and one can define a duality operator for the horizontal 2-forms as
\begin{eqnarray}
 \omega\to *_{\sreeb}\omega=\iota_{\sreeb}*\omega~,~~~\omega\in\Omega_H^2(M)~.\label{QZhorizontal_dual}\end{eqnarray}
The following relations are quite useful
\begin{eqnarray}
 \kappa\wedge *\omega=(-1)^{p-1}*\iota_{\sreeb}\omega~,~~~\iota_{\sreeb}*\omega=(-1)^p*(\kappa\omega)~,~~\omega\in\Omega^p(M)~.\label{QZsigns}
 \end{eqnarray}

\noindent $\bullet$ K-contact structure.
If ${\reeb}$ is a Killing vector field with respect to
$g$ of (\ref{QZstd_metric}), then $(\kappa,{\reeb},J)$ gives a K-contact structure, the Killing condition is equivalent to
\begin{eqnarray}
\nabla_X{\reeb}=JX~,~~~\forall X\in TM~.\label{QZReeb_J}
\end{eqnarray}

\noindent $\bullet$ Sasaki manifolds.
From $M$, one can construct a manifold $C(M)$ which is a cone over $M$ with metric, symplectic and almost complex structures
\begin{eqnarray}
&& C(M)=\QZBB{R}^{>0}\times M~,\nn\\
&&{\QZcal G}=dt^2+t^2g~,\label{QZcone_metric}\\
&&\omega=d(t^2\kappa)~,\nn\\
&&{\QZcal J}=2\omega^{-1}{\QZcal G}~.\nn\end{eqnarray}
A Sasaki manifold is a K-contact manifold such that $(C(M),{\QZcal G},\omega,{\QZcal J})$ is K\"ahler. The complex structure is written explicitly as
\begin{eqnarray}
 {\QZcal J}=J+t^{-1}\reeb\otimes dt-t\partial_t\otimes \kappa~,\nn
 \end{eqnarray}
it is easy to check ${\QZcal J}^2=-1_6$. The vector field
\begin{eqnarray}
\ep=t\frac{\partial}{\partial t}\label{QZhomothetic}
\end{eqnarray}
generates a scaling along the $t$-direction and
is called the \emph{homothetic} vector field. It is clear that
\begin{eqnarray}
{\QZcal J}(\ep)=\reeb~.\label{QZep_reeb}
\end{eqnarray}

The K\"ahler condition is equivalent to the covariant constancy of ${\QZcal J}$ with respect to  the Levi-Civita connection. Thus a K-contact manifold is Sasaki iff $J$ satisfies the
integrability condition
\begin{eqnarray}
&&\QZbra Z,(\nabla_XJ)Y\QZket=-\kappa(Z)\QZbra X,Y\QZket+\QZbra Z,X\QZket\kappa(Y)\label{QZintegrability}~,
\end{eqnarray}
where $\QZbra-,-\QZket$ is the inner product using the metric
\begin{eqnarray}
\QZbra A,B\QZket=A_{i_1\cdots i_p}B_{j_1\cdots j_p}g^{i_1j_1}\cdots  g^{i_pj_p}~.\label{QZinner_product}
\end{eqnarray}

From (\ref{QZintegrability}) one can derive a wealth of relations, some of which will be needed later. Define first some more projectors
\begin{eqnarray}
 (P_\pm)_p^{~q}=\frac12(P\pm iJ)^{~q}_p~,\label{QZprojectors}
 \end{eqnarray}
where $P$ is the projection to the horizontal component of a vector or a form, with its indices written out $P^r_s=\delta^r_s-\reeb^r\reeb_s$. The two projectors project to the horizontal $(1,0)$ or $(0,1)$ component
 with respect to the complex structure $J$. Keeping in mind the K\"ahler property of the cone $C(M)$ will lead to the vanishing of (0,2) and (2,0) components of the curvature tensor, which translates in 5d as
\begin{eqnarray}
&&R_{mnpq}-J_p^{~u}J_q^{~v}R_{mnuv}=g_{p[m}g_{n]q}+J_{p[m}J_{n]q}~,\nn\\
&&R_{mnpq}\reeb^p=-g_{q[m}\reeb_{n]}~.\label{QZproperty_curvature}
\end{eqnarray}
The first equation says that
\begin{eqnarray}
 (P_-)_p^{~s}(P_-)_q^{~t}R_{mnst}=(P_-)_{pm}(P_-)_{qn}-(m\leftrightarrow n)~,\label{QZCurvature_2_0}
 \end{eqnarray}
i.e. the (0,2) component of the curvature, though not vanishing, can be written as something elementary.

It it useful to think of a Sasaki-manifold as an odd-dimensional analogue of a K\"ahler manifold. In fact, not only is the cone K\"ahler, there is also a K\"ahler structure transverse to the Reeb foliation (see \cite{QZ2010arXiv1004.2461S} or section 7 of \cite{QZBoyerGalicki}). One can develop a transverse Dolbeault or even the appropriate  Hodge theory. We define an operator
\begin{eqnarray}
 \bar\partial_H:~\Omega_H^{p,q}\to\Omega_H^{p,q+1}~,~~~\bar\partial_H=dx^r(P_-)_r^{~s}\nabla_s+iq\kappa~.\label{QZbar_partial_H}
 \end{eqnarray}
It is a differential that sends $\Omega_H^{p,q}\to \Omega_H^{p,q+1}$. It is a bit lengthy but straightforward to check the claimed properties of this operator, so we suppress the proof, but one needs to make use of equations (\ref{QZReeb_J}) (\ref{QZintegrability}) and (\ref{QZproperty_curvature}).
It is also useful to transcribe the 6d Dolbeault operator in 5d language
\begin{eqnarray}
\bar\partial^6
=\frac{1}{2}(t^{-1}dt-i\kappa)(L_{\ep}+iL_{\sreeb})+(P_-)_{pq}dx^pdx^q\iota_{\ep}+\bar\partial_H\label{QZdolbeault_6_5}~.
\end{eqnarray}

\noindent $\bullet$ Sasaki-Einstein manifolds. If the cone metric is in addition Ricci-flat, i.e. the cone is Calabi-Yau, then $M$ is said to be Sasaki-Einstein (SE), which is the central player in this review. The Ricci flatness of the cone is equivalent to 5d condition
\begin{eqnarray} R_{mn}=4g_{mn}~.\label{QZEinstein}\end{eqnarray}

The CY property implies that there is a nowhere vanishing section of (3,0) forms on the cone, let us pick a harmonic representative $\Omega$, i.e. $\bar\partial^6\Omega=0$ (since the cone is neither compact nor smooth, one needs to construct the harmonic representative explicitly).
From $\Omega$ we define $\varrho=\iota_{\ep}\Omega=-i\iota_{\sreeb}\Omega$, the restriction of $\varrho$ to the surface $t=1$, i.e. to $M$ will be a nowhere vanishing section of $\Omega_H^{2,0}(M)$ already appearing in sec.\ref{QZsec_Hmavt}. From $\bar\partial^6\Omega=0$ and the relation (\ref{QZdolbeault_6_5})
\begin{eqnarray}
 0=\bar\partial^6\Omega=\frac{i}{2}t^{-1}dt\kappa(L_{\ep-i\sreeb}\varrho)-\frac12(t^{-1}dt+i\kappa)\bar\partial_H\varrho~.\nn
 \end{eqnarray}
Since $L_{\ep-i\sreeb}\varrho\in\Omega_H^{2,0}$ and $\bar\partial_H\varrho\in\Omega_H^{2,1}$, we have
\begin{eqnarray}
 L_{\ep-i\sreeb}\varrho=0=\bar\partial_H\varrho~.\nn
\end{eqnarray}
From the last equation we also get
\begin{eqnarray}
\nabla^{\dag}\varrho=0~.\nn
\end{eqnarray}

\subsection{Toric Sasaki manifolds} \label{QZsec_TSm}

This section presents the construction of  examples for the manifolds discussed in the previous subsection. To construct Sasaki-manifolds, it is easier to start from its K\"ahler cone, which can be obtained through K\"ahler reduction from a flat space.

Consider $\QZBB{C}^4$ with the standard K\"ahler structure. Let ${\tt e}_i,~i=1,\cdots,4$ be four $U(1)$'s that rotate the phase of each $\QZBB{C}$ factor. The $U(1)$ actions are Hamiltonian with moment map
\begin{eqnarray}
 \vec\mu=\frac12(|z_1|^2,\cdots,|z_4|^2)~.\nn
 \end{eqnarray}
Let $U(1)_T=T^i{\tt e}_i$ be a particular combination of these $U(1)$'s, we can assume that $T$ is primitive, i.e. the four components of $T$ have greatest common divisor 1.
The action of $U(1)_T$ is has moment map
\begin{eqnarray}
 \mu_T=\vec T\cdotp\vec\mu=\frac12\sum_{i=1}^4\,|z_i|^2T^i~,\nn
 \end{eqnarray}
suppose that the four components of $T$ are not all positive or negative, then
$\mu_T^{-1}(0)$ is non-trivial. Let
\begin{eqnarray}
C(M)=\mu^{-1}_T(0)/U(1)_T\nn
\end{eqnarray}
be the K\"ahler reduction of $\QZBB{C}^4$. Since $\mu_T^{-1}(0)$ is invariant under the simultaneous scaling $z_i\to \lambda z_i$, $\lambda\in\QZBB{R}^{\times}$, hence $\mu^{-1}_T(0)$ and $C(M)$ have the structure of a cone. Note that the action of $U(1)_T$ on $\mu_T^{-1}(0)$ is not free, so $C(M)$ will always be singular.

From $C(M)$, we can get to $M$ by imposing a constraint to fix the scaling freedom above, by intersecting the $C_{\mu}(M)$ with a hyper-surface. %provided that the intersection, which we call 'base', is a convex polygon. Then one can understand the geometry of $M$ also as a torus fibration over this polygon, where various tori degenerate at its boundary, in the same fashion as discussed above.
%Care must be taken to ensure that $M$ obtained as a torus fibration is smooth.
We pick a 4-vector (not necessarily integer) $\vec\omega$ and consider the surface
\begin{eqnarray}
H_{\omega}=\{z_i\in\QZBB{C}|\sum_{i=1}^4\omega_i|z_i|^2=1\}~.\label{QZHP_omega}
\end{eqnarray}
The $U(1)_T$ action on the intersection $H_{\omega}\cap \mu_T^{-1}(0)$
can be free if $T,\omega$ are appropriately chosen. As an example, let $T=[p+q,p-q,-p,-p]$, with $p>q>1$ and $\delta(p,q)=1$. Also choose $\omega_i>0,~i=1,\cdots,4$, then the intersection is topologically $S^3\times S^3$. The loci where $U(1)_T$ action is non-free is at $z_1=z_2=0$ or $z_3=z_4=0$, both of which are excluded by the intersection. With a free action secured, the quotient
\begin{eqnarray} M=H_{\omega}\cap \mu_T^{-1}(0)/U(1)_T\nn\end{eqnarray}
is a smooth 5d manifold. In fact, the $U(1)$ determined by $\sum_{i=1}^4\omega_i{\tt e}_i$ serves as the Reeb vector field on $M$.

We can give a more intrinsic description of $C(M)$ and $M$. Out of the four $U(1)$'s acting on $z_i$, there will be only three independent $U(1)$'s left after the K\"ahler reduction, let us pick a basis $e^a=1,2,3$ for them. An explicit such basis can be chosen as follows.
With our assumption on the primitiveness of $T$, we can find a $4\times 4$ matrix $A\in SL(4,\QZBB{Z})$ with $T$ as the last column,
then the linear combinations
\begin{eqnarray}
e_a=\sum_{i=1}^4 {\tt e}_iA^i_a~,~~a=1,2,3\nn
\end{eqnarray}
give a basis of the three $U(1)$'s acting on $C(M)$. Denote the Hamiltonian of the three $U(1)$'s as $y_a,~a=1,2,3$. The $y_a$ is an explicit parametrisation of the hyperplane $\sum_i T^i|z_i|^2=0$.

Write now
\begin{eqnarray}
(A^{-1})^a_i=\left[\begin{array}{cccc} \vec v_1 & \vec v_2 & \vec v_3 & \vec v_4\\
\cdots & \cdots & \cdots & \cdots \end{array} \right]_{4\times 4}~,\nn\end{eqnarray}
where $\vec v_i$ are integer 3-vectors. On the hyperplane $\sum_i T^i|z_i|^2=0$, the $|z_i|^2$ are solved as
\begin{eqnarray}
0\leq\frac12|z_i|^2=\sum_ay_av^a_i~.\label{QZinequ_mu}
\end{eqnarray}
The inequalities (\ref{QZinequ_mu}) demarcates the domain of $\{y_a\}$ as being a polytope cone. This cone actually contains almost all information about the geometry, so we give it a name $C_{\mu}(M)$.
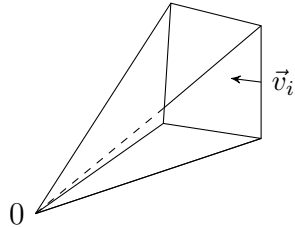
\begin{figure}[h]
\begin{center}
\begin{tikzpicture}[scale=1]
    \draw (0,0) node[left] {0} -- (3,1) -- (3,2.5) -- (1.8,2.8) -- (1.7,1.2) -- (3,1);
    \draw (0,0) -- (3,1);
    \draw[dashed] (0,0) -- (1.7,1.4);
    \draw (1.7,1.4) -- (3,2.5);
    \draw (0,0) -- (1.8,2.8);
    \draw (0,0) -- (1.7,1.2);
%    \node at (1.7,1.4) {$\times$};
    \draw[->] (3,1.75) node[right] {\small{$\vec v_i$}}-- (2.6,1.8);
\end{tikzpicture}\caption{The moment map cone $C_{\mu}(M)$}\label{QZfig_mom_map}
\end{center}
\end{figure}
Referring to fig.\ref{QZfig_mom_map}, the inward pointing normals are $\vec v_i,~i=1,\cdots,4$ (though the order may not be the same as how they appear in $A^{-1}$). A good way to visualize the geometry of $C(M)$ is that it is a torus fibration over $C_{\mu}(M)$. A generic fibre is $U(1)^3$, but the tori may degenerate at the boundaries of the cone. At the codim 1 faces, say face $1$, the $U(1)$ singled out by $\vec v_1$, i.e. $\sum_ae_av^a_i$ degenerates, while at the intersection of faces 1 and 2, two $U(1)$'s singled out by $\vec v_{1,2}$ degenerate, etc.

To complete our translation of the geometry of $M$ into that of $C_{\mu}(M)$, let $\vec \reeb$ be a 3-vector with components
\begin{eqnarray}
\reeb^a=\sum_{i=1}^4v_i^a\omega_i~.\label{QZreeb_omega}
\end{eqnarray}
The 3-vector $\vec\reeb$ gives a linear combination of $U(1)$'s: $\sum_{a=1}^3\reeb^ae_a$, this $U(1)$ is the Reeb vector field on $M$, now written in purely 6d terms. Due to the correspondence (\ref{QZreeb_omega}), we will call both $\vec\omega$, $\vec\reeb$ the Reeb vector (that they represent).

Furthermore the condition $H_{\omega}$ of (\ref{QZHP_omega}) translates to
\begin{eqnarray}
\vec\reeb\cdotp\vec y=\frac12~.\nn
\end{eqnarray}
The intersection of this hyper-plane with $C_{\mu}(M)$ is a compact polygon iff $\vec\reeb$ is within the dual cone of the cone $C_{\mu}(M)$, i.e. $\vec\reeb=\sum_{i=1}^4\lambda_i\vec v_i,~\lambda_i>0$.
This compact polygon is the base of $U(1)^3$ fibration, whose total space is the celebrated $Y^{p,q}$ manifold \cite{QZGauntlett:2004yd}. We have also an easy generalisation
\begin{QZexample}A close cousin of $Y^{p,q}$ is obtained by taking $T=[a,b,-c,-a-b+c]$ such that $a,b,a+b-c>0$, $\delta(a,c)=\delta(a,d)=\delta(b,c)=\delta(b,d)=1$, known as the $L^{a,b,c}$ space.

Then $T$ can be completed into an $SL(4,\QZBB{Z})$ matrix
\begin{eqnarray} A=\left[\begin{array}{cccc} 0 & m & 0 & a \\
                         0 & 0 & 1 & b \\
                         0 & n & 0 & -c\\
                         1 & -m-n & -1 & -a-b+c\end{array}\right]~,~~~mc+na=1~.\nn\end{eqnarray}
And its inverse is
\begin{eqnarray} A^{-1}=\left[\begin{array}{cccc} 1 & 1 & 1 & 1 \\
                         c & 0 & a & 0 \\
                         -bn & 1 & bm & 0\\
                         n & 0 & -m & 0 \end{array}\right]~,\nn\end{eqnarray}
and from the first three rows we read off the inward normals (in their right order)
\begin{eqnarray} \vec v_1=[1,c,-bn]~,~~\vec v_2=[1,a,bm]~,~~\vec v_3=[1,0,1]~,~~\vec v_4=[1,0,0]~.\label{QZex_Labc}\end{eqnarray}
\end{QZexample}
These are two of the few SE manifolds, for which we know the explicit metric \cite{QZGauntlett:2004yd} \cite{QZCvetic:2005ft}.

The same story above can be repeated starting from $\QZBB{C}^{\tt n+3}$ and a K\"ahler reduction with respect to
 $\tt n$-charges.
But we stress that one does not have to take the route of the K\"ahler reduction, rather one may start from the more fundamental object $C_{\mu}(M)$. For example, one can postulate a polytope cone $C\subset \QZBB{R}^n$, with inward pointing normals $\vec v_i,~i=1\cdots N$ (assumed to be primitive of course), then Lehman \cite{QZ2001math......7201L} showed that if at the intersection of $k$ ($k\leq n-1$) faces, the $k$ normals $\vec v_1,\cdots,\vec v_k$ satisfy
\begin{eqnarray}
\textrm{span}_{\QZBB{Z}}\QZbra\vec v_1,\cdots,\vec v_k\QZket=\textrm{span}_{\QZBB{R}}\QZbra\vec v_1,\cdots,\vec v_k\QZket\cap\QZBB{Z}^n~,\nn\end{eqnarray}
then the cone gives rise to a smooth toric contact manifold. These conditions can be explicitly checked for the $Y^{p,q},~L^{a,b,c}$ cases above (the explicit normals and a more convenient criteria are given in \cite{QZQiu:2014oqa}).

\smallskip

\noindent {\it Toric Sasaki-Einstein manifolds}. By definition if $M$ is toric SE, then its metric cone is CY and then in the K\"ahler reduction construction of $C(M)$ the charges of $U(1)_T$ must sum to zero. This has a very simple implication when translated into the cone language: there exists a primitive $\vec\xi\in\QZBB{Z}^3$ such that
\begin{eqnarray}
\vec\xi\cdotp\vec v_i=1~,~~~i=1,\cdots,\tt n\label{QZ1-gorenstein}\end{eqnarray}
known as the 1-Gorenstein condition.
The proof of this fact is not difficult and is left to the reader.
Referring to the example above (\ref{QZex_Labc}), all $\vec v_i$ has its first component equal to 1, and so one chooses simply $\vec\xi=[1,0,0]$.

Since $C(M)$ has flat canonical bundle, and if it is also simply connected, we will have a nowhere vanishing section $\Omega$, whose contraction with $\ep$ gives the $\varrho$ in the previous section.

\smallskip

\noindent {\it Deformations}.

So far we have given the Reeb vector field, but not quite the rest of the contact structures.
Let us denote by ${\QZcal J}_0$ the standard complex structure on $\QZBB{C}^4$, then it descends through the K\"aher reduction to a complex structure on $C(M)$.
%The 4-vector $\vec \omega$ gives a linear combination of $U(1)$: $\sum_{i=1}^4\omega_i{\tt e}_i$ %acting on $M$.
Let also $\ep$ be the homothetic vector that scales all $z_i$, it is easy to observe
\begin{eqnarray}
 {\QZcal J}_0(\ep)={\QZcal J}_0\sum_{i=1}^4(z_i\partial_{z^i}+c.c)=i\sum_{i=1}^4(z^i\partial_{z^i}-c.c)=\sum_{i=1}^4{\tt e}_i~.\nn
 \end{eqnarray}
Comparing this with (\ref{QZep_reeb}), we have the Reeb $\vec\omega=[1,1,1,1]$, or using  (\ref{QZreeb_omega})
\begin{eqnarray}
 \vec\reeb_0=\sum_{i=1}^4\vec v_i~,\nn
 \end{eqnarray}
which is certainly within the dual cone. We call this the standard Reeb and the corresponding complex structure the standard complex structure. But to obtain general Reeb vector fields, one needs to deform ${\QZcal J}$, which can be done in a very transparent manner in the toric setting. As these deformations are reflected in the partition function, so using susy gauge theory as a means to study these geometries is an interesting enterprise.

Since we are interested in toric deformations, it is convenient to to use $y^i=|z_i|^2/2,~\theta_i=\arg z_i$ as coordinates of $\QZBB{C}^{\tt n+3}$, we just take $\tt n=1$ as an illustration. The material here is taken from the marvelous paper \cite{QZMartelli:2005tp}.
One can write a metric
\begin{eqnarray}
 &&{\QZcal G}=G_{ij}dy^idy^j+G^{ij}d\theta_i d\theta_j~,~~~G_{ij}=\partial_{y^i}\partial_{y^j}G~,~~G^{ij}=G^{-1}_{ij}~,\nn\\
&&G=\frac12\sum_{i=1}^4y^i\log y^i+\frac12(\vec\omega\cdotp\vec y)\log(\vec\omega\cdotp\vec y)-\frac12(\vec\omega_0\cdotp\vec y)\log(\vec\omega_0\cdotp\vec y)~,\nn
\end{eqnarray}
where $\vec\omega_0=[1,1,1,1]$. Even though ${\QZcal G}$ appear to have singularities at $y^i=0$, they are only coordinate singularities, in fact, when $y^1\sim 0$
\begin{eqnarray}
 {\QZcal G}\sim \frac12\frac{dy^1dy^1}{y^1}+2y^1d\theta_1 d\theta_1+\cdots~,\nn
\end{eqnarray}
and is perfectly smooth at $y^1=0$ after reverting to Cartesian coordinates. We need not worry too much about the positivity of ${\QZcal G}$, as ${\QZcal G}$ is certainly so when $\omega=\omega_0$, and thus there is an open neighbourhood round $\omega=\omega_0$ in which positivity is secured.

The complex structure is
\begin{eqnarray}
{\QZcal J}=-G_{ij}dy^i\otimes\partial_{\theta_j}+G^{ij}d\theta_i\otimes \partial_{y^j}~.\label{QZcomplex_str}
\end{eqnarray}
To see that it is integrable, consider the (0,1)-forms $d\theta_i+iG_{ij}dy^j$, since
$d(d\theta_i+iG_{ij}dy^j)=iG_{ij,k}dy^k\wedge dy^j=0$, the distribution $\cap_i\ker (d\theta_i+iG_{ij}dy^j)$ is integrable, and ${\QZcal J}$ is integrable. In fact, if one works out explicitly the Levi-Civita connection, then ${\QZcal J}$ is covariantly constant, i.e. we have a  K\"ahler structure $\QZBB{C}^4$.

Since all structures here descend through the K\"ahler reduction, we have a deformed K\"ahler structure on the cone $C(M)$. In particular
\begin{eqnarray}
J(\ep)=2J(y^i\partial_i)=2G_{ij}y^j\partial_{\theta_i}=\sum\omega^i\partial_{\theta_i}~,\nn
\end{eqnarray}
comparing with (\ref{QZep_reeb}), and using the correspondence (\ref{QZreeb_omega}) of two presentations of the Reeb before and after the K\"ahler reduction, one reaches the desired Reeb vector field.

So far we have held the symplectic structure of the cone fixed and deformed the complex structure, and hence the use of the symplectic coordinates $y,\,\theta$. In the subsequent discussion of deformation of the symplectic structure (which we will not need in this paper, since the partition function only depends on the complex moduli), one must switch to the complex coordinates. From the explicit complex structure above, one can identify the (1,0)-forms
\begin{eqnarray}
(1+i{\QZcal J})d\theta_i=d\theta_i-iG_{ij}dy^j=d(\theta_i-iG_i)~.\nn
\end{eqnarray}
Thus we let the complex coordinates be
\begin{eqnarray}
 z_i=\exp{(G_i+i\theta_i)}~.\nn
 \end{eqnarray}
For example letting $\reeb=\reeb_0$ in (\ref{QZcomplex_str}), one can check that $z_i$ is a constant multiple of the standard complex coordinates of $\QZBB{C}^{\tt 4}$.

With the explicit complex coordinates one can construct a harmonic representative of the holomorphic volume form $\Omega$ from the K\"ahler reduction picture. Take $\Omega_0=dz_1\wedge\cdots\wedge dz_4$, which is smooth and is also invariant under (the complexified) $U(1)_T$ action since the charges of $U(1)_T$ sums to zero. Now let $\Omega=\iota_T\Omega_0$,
where $T_i$ are the vector fields induced by the $U(1)_T$ action. Thus $\Omega$ is a basic form with respect to
  $U(1)_T$ and thus descends through the K\"ahler reduction and becomes the holomorphic volume form of $C(M)$. Moreover as $\Omega_0$ is annihilated by $\bar\partial$ and so will be $\Omega$, thus $\Omega$ is a harmonic representative of $H^{0,3}(C(M))$. It is is easily checked that $\Omega$ scales under $\ep$ with weight $\sum_k\omega_k$, subsequently has weight $i\sum_k\omega_k$ under $L_{\sreeb}$.
This is how one can obtain the weight of $\varrho$ under $L_{\sreeb}$.

\documentfinish